\documentclass
[aps,prd,preprint,tightenlines,superscriptaddress,nofootinbib]{revtex4}
\usepackage{amsmath,amssymb}
\usepackage{graphicx,tabularx}
\usepackage{amsmath}
\usepackage{amsfonts}
\usepackage{amssymb}
\usepackage{graphicx}
\def\bq{\begin{equation}}
\def\eq{\end{equation}}
\def\ba{\begin{eqnarray}}
\def\ea{\end{eqnarray}}
\newcommand{\sla}[1]{/\!\!\!#1}

\begin{document}
\date{\today}
\title{Standard Model Top Quark Asymmetry at the Fermilab Tevatron}
\author{M.~T. Bowen}
\email{mtb6@u.washington.edu}
\author{S.~D.~Ellis}
\email{ellis@phys.washington.edu}
\affiliation{Dept. of Physics, University of Washington, Seattle, WA 98195}
\author{D.~Rainwater}
\email{rain@pas.rochester.edu}
\affiliation{Dept. of Physics and Astronomy, University of Rochester, Rochester, NY 14627}

\begin{abstract}
Top quark pair production at proton-antiproton colliders is known to
exhibit a forward-backward asymmetry due to higher-order QCD
effects. We explore how this asymmetry might be studied at the
Fermilab Tevatron, including how the asymmetry depends on the
kinematics of extra hard partons. We consider results for top quark
pair events with one and two additional hard jets. We further note
that a similar asymmetry, correlated with the presence of jets, arises
in specific models for parton showers in Monte Carlo simulations. We
conclude that the measurement of this asymmetry at the Tevatron will
be challenging, but important both for our understanding of QCD and
for our efforts to model it.
\end{abstract}

\maketitle

%%%%%%%%%%%%%%%%%%%%%%%%%%%%%%%%%%%%%%%%%%%%%%%%%%%%%%%%%%%%%%%%%%%%%%%%
%%%%%%%%%%%%%%%%%%%%%%%%%%%%%%%%%%%%%%%%%%%%%%%%%%%%%%%%%%%%%%%%%%%%%%%%

\section{Introduction}

The matter content of the highly successful Standard Model (SM) of
particle physics was generally considered to be fully revealed after
the discovery of the top quark in 1994~\cite{topdiscov}. While the
exact mechanism of electroweak symmetry breaking (EWSB) remains
undetermined~\cite{EWSB}, the unusually large top quark
mass~\cite{Azzi:2004rc,Abazov:2004cs}, of the same size as the EWSB
scale, suggests that the top quark may play a special role.  Taking a
closer experimental look at the top quark is therefore high on our
list of priorities -- not to mention that its quantum numbers (other
than mass) and couplings to other SM particles are only crudely known,
or not known at all~\cite{Chakraborty:2003iw}. Furthermore, the
production and subsequent decay of the top quark at current and future
collider experiments is a serious background to many new physics
searches. Thus it is essential to study top quark production in much
more detail, and eventually to be able to simulate it accurately in
Monte Carlo background studies.

The top quark sample that exists today, exclusively from Runs~I and II
of the Fermilab Tevatron, is rather small, a few hundred events. This
shall soon increase by an order of magnitude or more. The Tevatron
experiments CDF and D\O \ are beginning to study the amount of
additional jet activity in top quark pair events, and early
indications are that it is non-trivial~\cite{data-ttj}. There are also
hints in the data from Run~I that the kinematic distribution of the
top quark pairs' decay products exhibit a charge
asymmetry~\cite{vejcik}.

We reexamine QCD top quark pair production in $p\bar{p}$ collisions at
Tevatron Run~II energy, with special attention to the asymmetries in
the production process and to the correlation of the asymmetries with
extra radiation in top quark pair events. In fact, the presence of \
both the asymmetries and extra radiation have been theoretically
expected for some time, but neither the structure of the asymmetries
nor their correlations with extra radiation are fully reproduced by
the standard tools used in experimental analysis. We make the case
here that it is now time to study the asymmetry and its correlation
with the extra radiation in detail. In Sec.~\ref{sec:prod} we review
inclusive QCD top quark pair production at hadron colliders, including
known next-to-leading order (NLO) effects.  Previous work had
identified an overall asymmetry for top quark pairs at leading order
(LO)~\cite{Halzen:1987xd} and NLO~\cite{Kuhn}, although it did not
identify all the kinematic features present. We present updated
results for the asymmetry at NLO, for both inclusive and exclusive
samples. In Sec.~\ref{sec:tt1j} we present new results for the real
emission component of inclusive $t\bar{t}$ production, where the
asymmetry is shown as a differential distribution as a function of the
kinematics of the extra hard radiation. We provide an analysis of the
likely statistical uncertainties in a variety of scenarios with
various luminosities, kinematic cuts and tagging efficiencies. \ We do
not consider the questions of systematics and background rates in any
detail. \ In Sections~\ref{sec:excl0j} and~\ref{sec:tt2j} we study the
asymmetry for $t\bar{t}$ production in the zero jet exclusive sample
(additional jet activity is vetoed) and double jet sample,
respectively.  Section~\ref{sec:PSMC} provides a discussion of the
level of asymmetry that can be found in parton-shower Monte Carlo
(PSMC) results where the NLO perturbative (coherent) asymmetry
effects, as described in the previous sections, are not
present. Finally we review our results and conclude in Section
~\ref{sec:conc}.

%%%%%%%%%%%%%%%%%%%%%%%%%%%%%%%%%%%%%%%%%%%%%%%%%%%%%%%%%%%%%%%%%%%%%%%%

\section{Inclusive QCD top quark pair production \label{sec:prod}}

At LO, $t\bar{t}$ production is totally charge-conjugation symmetric
for both production mechanisms, quark- and gluon-fusion: $q\bar{q}\to
t\bar{t}$ and $gg\to t\bar{t}$. As a consequence, the angular
distributions of the $t$ and $\bar{t}$ are totally symmetric for
$p\bar{p}$ collisions. However, at higher orders in $\alpha_{s}$, this
no longer remains true for all subprocesses.

As was pointed out almost two decades ago~\cite{Halzen:1987xd}, not
all processes involving additional partons are symmetric under charge
conjugation with respect to the incoming proton and anti-proton
beams. The process $gg\to t\bar{t}g$ is, but the processes
$q\bar{q}\to t\bar {t}g$ and $qg\to t\bar{t}q$ are not. Processes
involving initial state valence quarks will therefore exhibit a
forward-backward asymmetry. This is caused by interference between
initial- and final-state gluon emission (or its crossed process),
analogous to what happens in QED, \textit{e.g.}, the forward-backward
asymmetry in $e^{+}e^{-}\to\mu^{+}\mu^{-}\gamma$ or other heavy
fermion pairs~\cite{QED-analogue}. Because $t\bar{t}$ production at
the Tevatron is dominated at the $90\%$ level by $q\bar{q}$
annihilation, we can expect the $q\bar{q}$ subprocess' asymmetry to be
closely reflected in the total sample.

When the $p\bar{p}\to t\bar{t}$ cross section was calculated more
fully at NLO~\cite{tt-NLO}, an asymmetry due to virtual corrections
was noticed. It arises from an interference between the color-singlet
4-point (box) virtual correction and the Born term for the process
$q\bar {q}\to t\bar{t}$. Ref.~\cite{Kuhn} examined this asymmetry more
closely, although still inclusively and integrated over the phase
space of the additional parton. It was found that the virtual
contribution produces an asymmetry opposite in sign to and larger than
that of the real emission component. For the forward-backward
inclusive asymmetry defined in terms of the top quark rapidity
$y_{t}$\footnote{Here we use rapidity, $y$, instead of polar angle,
$\theta$, since rapidity is conventional for hadron collider
experiments. For massless particles, rapidity $y$ is equal to
pseudorapidity $\eta$.},
\bq\label{eq:asym}
A_{FB}^{t}\,=\,\frac{N_{t}(y_{t}>0)-N_{t}\,(y_{t}<0)}{N_{t}(y_{t}
>0)+N_{t}\,(y_{t}<0)}\;,
\eq
Ref.~\cite{Kuhn} calculated a value of $4-5\%$ for the Tevatron Run~I,
$\sqrt{s}=1.8\,${TeV}, and top quark mass $m_{t}=175\,${GeV}.

%%%%%%%%%%%%%%%%%%%%%%%%%%%%%%%%%%%%%%%%%%%%%%%%%%%%%%%%%%%%%%%%%%%%%%%%

\subsection{Recalculation of top quark asymmetry}

\label{sec:recalc}

\begin{figure}[th]
\begin{center}
\includegraphics[scale=0.83]{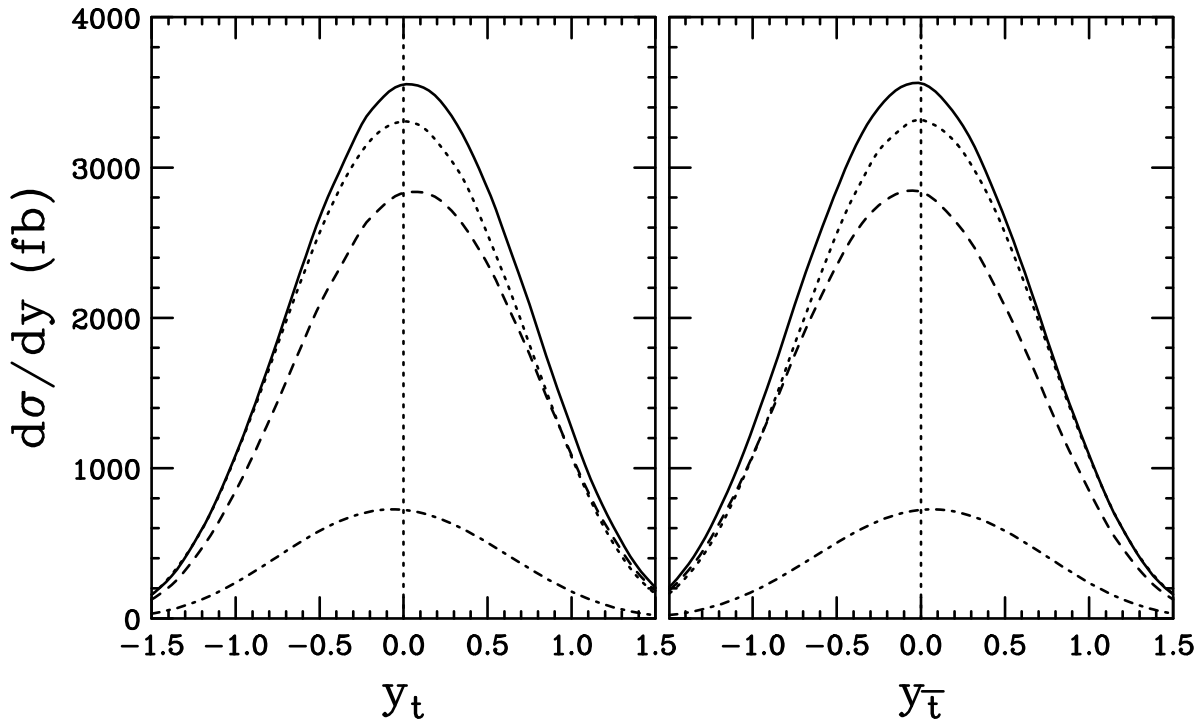} \vspace{-7mm}
\end{center}
\caption{Differential cross section distributions as a function of the
top (left) and anti-top (right) quark rapidities, produced in
$p\bar{p}$ collisions at the Tevatron in Run II,
$\sqrt{s}=1.96\;${TeV}. Shown are the LO $t\bar{t}$ inclusive
(dotted), $t\bar{t}$ NLO inclusive (solid), LO $t\bar {t}j$ inclusive
(dot-dashed) and $t\bar{t}0j$ exclusive (dashed) predictions.  We
define the LO $t\bar{t}j$ inclusive rate as that where the additional
final-state parton has $p_{T}(j)>20$~GeV and $|\eta(j)|<3$.  The
$t\bar{t}0j$ exclusive rate is then the NLO inclusive rate minus the
LO $t\bar{t}j$ inclusive rate.}
\label{fig:asym.NLO}
\end{figure}

To provide a connection to the previous work outlined above, we have
recalculated the top and anti-top quark distributions for Tevatron
Run~II, $\sqrt{s}=1.96\,${TeV}, using the NLO Monte Carlo code
\textsc{mcfm} ~\cite{MCFM} with the updated top quark mass,
$m_{t}=178\,${GeV}, and structure functions of
CTEQ6~\cite{Pumplin:2002vw} (set L1 for LO, M for NLO).  We show our
results in Fig.~\ref{fig:asym.NLO} for the LO and NLO inclusive
calculations, and also for the $t\bar{t}0j$ exclusive calculations. We
define this last quantity as the NLO $t\bar{t}$ inclusive rate minus
the LO $t\bar{t}j$ inclusive rate above some cutoff,
$p_{T}(j)>p_{T}(j,min)$ and inside some rapidity region
$|\eta(j)|<\eta(j,max)$. (Note that inclusive NLO $t\bar{t}$ and LO
$t\bar{t}j$ are the same order in $\alpha_{s}$.) Thus we have
\bq
\sigma_{t\bar{t}0j}^{\mathrm{excl}}\;=\;\sigma_{t\bar{t}}^{\mathrm{NLO}
}\,-\,\sigma_{t\bar{t}j}^{\mathrm{LO}}\biggl(p_{T}(j)>p_{T}(j,min),|\eta
(j)|<\eta(j,max)\biggr)\;. \label{eq:0j-excl}
\eq
The $t\bar{t}0j$ exclusive rate corresponds to the total $t\bar{t}+X$
rate with a jet veto applied above some $p_{T}$ inside some fiducial
region of rapidity in the detector. Here we use the cutoffs
\bq
p_{T}(j)>20\,\text{{GeV}}\;,\qquad|\eta(j)|<3.0\;.\nonumber
\eq
This fiducial region is based on the Run~II capabilities of the CDF
and D\O\ detectors. It is important to note that, in order to evaluate
the perturbative $t\bar{t}j$ asymmetry described above, the
calculation \textit{must} employ the exact matrix elements. Since
parton-shower Monte Carlo tools such as \textsc{pythia}~\cite{pythia},
\textsc{herwig}~\cite{herwig} and \textsc{sherpa}~\cite{sherpa} do
not include the full interference effects in the parton radiation that
produce the asymmetry, they cannot be used to predict the
asymmetry. However, as we will discuss later, the soft radiation
models used in the Monte Carlos are capable of producing certain kinds
of asymmetries, which must be understood in order to accurately
analyze any asymmetries that may be in the experimental data. In our
perturbative calculations the minimum $p_{T}$ kinematic cut protects
us from the soft singularity present in the $t\bar{t}g$ final
state. We verify that this is sufficient by checking that
$\sigma_{t\bar{t}j}\ll\sigma_{t\bar{t}}$.  For the kinematic cuts
defined above, \textsc{mcfm} gives $\sigma_{t\bar{t}}=5.9\,{pb}$ and
$\sigma_{t\bar{t}j}=1.1\,{pb}$. Their ratio is only 0.19, indicating
that the calculation is perturbative. However, because
$\sigma_{t\bar{t}j}$ is calculated only at LO, there is a considerable
residual uncertainty of probably a factor of two at the Tevatron
energy. This means there is also a larger uncertainty in the
$t\bar{t}0j$ exclusive rate compared to the NLO inclusive rate, since
the former is effectively calculated at LO. For example, we choose the
top quark mass, $m_{t}$, as the factorization and renormalization
scales in our benchmark cross section. This is conservative because it
lies at the low end of our estimates for the LO $t\bar{t}j$
rate. However, if we choose scales that take into account the softness
of the additional radiation parton, \textit{e.g.}, $p_{T}(j)$, the
cross section can be twice as large as our benchmark point. Even in
this case we would still regard our result as perturbative. For a
final analysis from Run~II data, knowing the NLO $t\bar{t}j$ rate will
probably become necessary.  Fortunately, this calculation is
underway~\cite{ttj-NLO}.

The structure of the asymmetry described above can be seen in
Fig.~\ref{fig:asym.NLO}. \ First note that the LO $t\bar{t}$ inclusive
cross section (dotted curve) is symmetric about $y=0$ in both panels,
\textit{i.e}., for both $t$ and $\bar{t}$. \ Comparison of this
symmetric curve with the (solid curve) NLO inclusive result indicates
that the latter curve is (slightly) shifted to larger $y$ for $t$
(left panel) and smaller $y$ for $\bar{t}$ (right panel) corresponding
to a positive asymmetry. \ The $t\bar{t}0j$ exclusive cross section
(dashed curve) exhibits a shift that is similar in terms of magnitude
and direction. \ On the other hand, the LO $t\bar{t}j$ inclusive cross
section (dot-dashed curve) is shifted in the opposite direction in
each panel yielding a negative asymmetry. \ Thus we conclude that the
real emission corrections tend to push the top quark backward,
opposite to the proton beam direction, while the virtual corrections
push it forward, in the direction of the proton beam. The virtual
corrections are larger as indicated by the qualitative agreement
between the overall NLO inclusive distribution (solid curve) and the
$t\bar{t}0j$ exclusive distribution (dashed curve). \ The $y_{t}$ and
$y_{\bar{t}}$ distributions are mirror images of each other, as
required by $CP$-invariance of QCD.

\begin{table}[th]
\begin{center}
\begin{tabular}
[c]{|l@{\hspace{0.5ex}}c|}\hline
\multicolumn{1}{|c|}{process} & \multicolumn{1}{c|}{$A^{t}_{FB}$}\\\hline
LO $t\bar{t}$ inclusive & 0 \\
NLO $t\bar{t}$ inclusive & \phantom{-}3.8 \\
LO $t\bar{t}j$ inclusive & -6.9\\
$t\bar{t}0j$ exclusive & \phantom{-}6.4 \\\hline
\end{tabular}
\vspace{-5mm}
\end{center}
\caption{Top quark forward-backward asymmetry in 
$p\bar{p}\to t\bar{t}$ production at Tevatron Run~II,
$\sqrt{s}=1.96\,{ TeV}$, for top quark mass $m_{t}=178\,{ GeV}$. The
$1j$ inclusive and $t\bar{t}0j$ exclusive rates are defined by
$p_{T}(j)>20\,{ GeV}$ and
$|\eta(j)|<3.0$. $A^{\bar{t}}_{FB}=-A^{t}_{FB}$ by
$CP$-invariance. The cross sections were calculated with
\textsc{mcfm}~\cite{MCFM}).}
\label{tab:asym.NLO}
\end{table}

Table~\ref{tab:asym.NLO} shows our numerical results for $A_{FB}^{t}$
that arise from integrating over the distributions in
Fig. \ref{fig:asym.NLO}. \ Note that, as expected, these estimates of
the magnitude of the $t\bar{t}$ asymmetries depend on the choice of
factorization and renormalization scales.  For example, with both
scales set to $m_t/2$, $m_t$, and $2m_t$, the NLO inclusive $t\bar{t}$
asymmetries are $5.5\%$, $3.8\%$, $2.8\%$, respectively.  In this
paper, we examine all asymmetries with both scales set to $m_t$.
\ Since the strong interaction is separately $C$ and $P$\ symmetric 
and the initial state in the $p\bar{p}$ process is a $CP$ eigenstate,
overall $CP$-invariance requires that
$A_{FB}^{\bar{t}}=-A_{FB}^{t}$. \ At the same time, since the initial
state is not an eigenstate of $C$ (or $P$) alone, we can have
$A_{FB}^{\bar{t}}\neq A_{FB}^{t}\neq0$. \ As expected from the
distributions, the LO asymmetry vanishes, the NLO inclusive and
$t\bar{t}0j$ exclusive asymmetries are positive, and the asymmetry
result for the $t\bar {t}j$ inclusive distribution is negative. \ When
considering these results, it is important to recall that the
magnitudes (but not the signs) of the last two entries in the table
depend on the $p_{T}$ cut defining the jets, which in this case has
the value $p_{T}(j)>20\,${GeV}. \ NLO final states with extra real
emission above the cut are included in the LO $t\bar{t}j$ inclusive
sample and exhibit a negative asymmetry. \ The corresponding real
emission configurations below the cut are included in the $t\bar{t}0j$
exclusive result, where they partially cancel the virtual
contributions with a positive asymmetry reducing the net asymmetry in
this term. Thus we can vary the magnitudes of the last two entries in
the table up and down together by varying the jet-defining $p_{T}$ cut
(larger magnitudes for a smaller cut and vice versa). Our numerical
results differ slightly from those of Ref.~\cite{Kuhn} due partly to
the higher energy, but also due to the different PDF set used here.

The obvious question to ask is, can Tevatron Run~II measure the
various inclusive and exclusive asymmetries predicted by perturbative
QCD? Observing the top quark asymmetry in a collider experiment such
as Tevatron Run~II is complicated by the difficulty in accurately
reconstructing the underlying parton kinematics from the measurements
of jets, compared to leptons. Because top quarks are produced at the
Tevatron with transverse momentum typically smaller than their mass,
their decay products are distributed over a large solid
angle. Mismeasurement of individual jets thus results in a
non-negligible uncertainty in the top quark direction. One should then
look instead at the decay products, since any forward-backward top
quark asymmetry will manifest itself in the daughter particles as
well, albeit with possibly different magnitudes. The $b$ jets are
tagged, but their charge is extremely difficult to determine, so they
are not good candidates. Nor are hadronically-decaying $W$ bosons,
again because of jet mismeasurement issues, but also because
light-flavor jet charges cannot be determined at all. Instead we
propose to use the asymmetry of the final state leptons (electron or
muon) from $W$ decays, since the the direction of leptons is extremely
well-measured.

Top quark pairs decay to two possible final states containing a
lepton: both top quarks can decay to $b\ell\nu$, where $\ell=e,\mu$,
which we call the
\textquotedblleft dilepton\textquotedblright\ final state; or one 
decays to $b\ell\nu$ while the other decays to $bjj$, which we call
the \textquotedblleft lepton+jets\textquotedblright\ final state. The
cleanest sample experimentally is the dilepton sample, although it
occurs at only about 1/6 the rate of the lepton+jets channel. Assuming
both $b$ jets are tagged, the dilepton sample has practically no
background from other SM processes, while the lepton+jets channel has
a signal to background (S:B) ratio of about 6:1~\cite{SBratio}. The
dilepton channel has the added advantage that there is no confusion
between the additional radiated hard jet in the event and jets from
the $W$ boson decay, as in the lepton+jets sample. Since these events
include jets along with the leptons, we will find it informative to
examine the correlations between the lepton forward-backward asymmetry
and the kinematic properties of the jets.

To obtain a handle on the expected lepton asymmetries we calculate
using matrix elements for $p\bar{p}\to t\bar{t}$ production, including
decay spin correlations and treating the intermediate $W$ bosons
off-shell, as generated by \textsc{madgraph}~\cite{Stelzer:1994ta}. We
evaluate cross sections at $\sqrt{s}=2.0$~TeV and use factorization
and renormalization scale choices of $\mu_{f}=\mu_{r}=m_{t}$ (the
small dependence on the total energy and the
factorization/renormalization scales will not be an issue here).
Before top quark decays, the cross section is 4.87~pb, compared to
6.7-8.0~pb for the NLO+(N)NLL $t\bar{t}$ rate~\cite{tt-NLO+}. We will
normalize our results to the NLO rate for our inclusive results; using
the average of the two NLO results, 7.4~pb, the effective K-factor is
1.52.

To calculate cross sections for observable final states at the parton
level, we use our LO \textsc{madgraph} code normalized to the NLO
rate, then impose generic kinematic cuts on the outgoing particles
suitable for both D\O\ and CDF:
\begin{align}
\label{eq:cuts} & \sla{p}_{T}>20~\mathrm{GeV} \; ,\nonumber\\
&  p_{T}(b) > 15~\mathrm{GeV} \; , \qquad|\eta(b)| < 2.0 \; , \qquad\Delta
R(b,b) > 0.4 \; ,\nonumber\\
&  p_{T}(j) > 20~\mathrm{GeV} \; , \qquad|\eta(j)| < 2.0 \; , \qquad\Delta
R(j,j;j,b) > 0.4 \; ,\\
&  p_{T}(\ell) > 20~\mathrm{GeV} \; , \qquad|\eta(\ell)| < 1.1 \; , \quad\Delta
R(\ell,\ell) > 0.2 \; ,\quad\Delta R(\ell,j;\ell,b) > 0.4 \; .\nonumber
\end{align}
The cuts are very conservative compared to the CDF and D\O\
expectations for the majority of Run~II, but reflect the present level
of understanding of the detectors and are used in current top quark
analyses~\cite{demina,tipton}. For example, muons can easily be
identified out to a pseudorapidity of 2.0, and non-$b$ jets can be
identified for $p_{T}(j)>15$~GeV. For both the dilepton and
lepton+jets final states we discuss below, we consider the possibility
of increasing the statistics by loosening the cuts accordingly. We do
not attempt to simulate detector effects.

Regardless of the decay mode, we define the lepton asymmetry as
\bq
A_{FB}^{\ell}\,=\,\frac{N(\eta_{\ell}>0)-N\,(\eta_{\ell}<0)}
{N(\eta_{\ell}>0)+N\,(\eta_{\ell}<0)}\;, \label{eq:asym-ll}
\eq
where the lepton may be $\ell^{+}$ or $\ell^{-}$, depending on which
is visible (or both, in dilepton events, but $A_{FB}^{\ell^{+}}$ and
$A_{FB}^{\ell^{-}}$ are calculated separately). Once a cross section
with cuts is obtained, the NLO-normalized number of expected events
for a given luminosity and set of detector ID efficiencies is
distributed into forward and backward bins such that the asymmetry is
the same $3.8\%$ as that predicted at NLO by \textsc{mcfm} \textit{for
the top quarks}. It is important to point out, as we will see later
explicitly in the $t\bar{t}j$ Section~\ref{sec:tt1j}, that kinematic
cuts can alter the asymmetry for a given subsample. Hence our
procedure is only an approximation. However, as no NLO $t\bar{t}$
program yet has the capability to decay the top quarks with spin
correlations intact at NLO, our approximation is the best we can
accomplish at the moment. Hopefully it will be superseded in the near
future.

%%%%%%%%%%%%%%%%%%%%%%%%%%%%%%%%%%%%%%%%%%%%%%%%%%%%%%%%%%%%%%%%%%%%%%%%

\subsection{Dilepton final state \label{subsec:incl-ll}}

When both top quarks decay leptonically, the final state in the
detector is $b\bar{b}\ell^{+}\ell^{-}\sla{p}_{T}$, where $\ell=e,\mu$
and $\sla{p}_{T}$ is missing energy transverse to the beam axis,
resulting from the neutrinos which escape undetected. We assume that
both $b$ jets are tagged, rendering the backgrounds negligible. This
is thus the cleanest top quark pair sample in the data. The cross
section with dilepton kinematic cuts and NLO K-factor applied is
113~fb.

We begin with a very conservative baseline assumption of $20\%$ for
the double $b$-tagging efficiency, which we label $\epsilon_{2b}$. For
4~fb$^{-1}$ and straightforwardly combining the statistics of D\O\ and
CDF (\textit{i.e.}, a factor of 2), we estimate about 180 events in
the dilepton sample with two $b$ tags summed over $\ell=e,\mu$ . The
statistical uncertainty on an asymmetry is given in
Ref.~\cite{Lyons:1986em}:
\bq
\delta A_{FB}^{\ell}\;=\frac{2\sqrt{N_{F}^{\ell}N_{B}^{\ell}}}{\sqrt{\left(
N_{F}^{\ell}+N_{B}^{\ell}\right)  ^{3}}}=\;\frac{1-\left(  A_{FB}^{\ell
}\right)  ^{2}}{2}\sqrt{\frac{1}{N_{F}^{\ell}}+\frac{1}{N_{B}^{\ell}}}
\eq
where $N_{F}^{\ell}(N_{B}^{\ell})$ is the number of observed events
with the lepton $\ell$ forward (backward). \ Taking $CP$ as a good
symmetry we have $A_{FB}^{\ell^{+}}=-A_{FB}^{\ell^{-}}$
($N_{F}^{\ell^{+}}=N_{B}^{\ell^{-}}$,
$N_{B}^{\ell^{+}}=N_{F}^{\ell^{-}}$) and we can combine the statistics
of the $\ell^{+}$ and $\ell^{-}$ channels. For the dilepton sample
this means that each event contributes twice. \ Combining the two
charges also yields a factor $\sqrt{2}$ improvement in overall
significance of the asymmetry determination over the single sign
analysis. We summarize the dilepton channel results in
Table~\ref{tab:delA-tt}. We see that the initial numbers for the
dilepton channel yield an uncertainty in the asymmetry measurement of
$\delta A_{FB}^{\ell}/A_{FB}^{\ell}\sim1.4$, providing only a
$0.7\sigma$ effect with the summed sign data. We conclude that it will
be challenging to use the dilepton channel due to poor statistics, at
least with our conservative assumptions. However, there are many
opportunities for increasing the sample size.

Here we explore various prospects for improving the sample size coming
from multiple sources, of varying likelihood. \ The results are
illustrated in Table \ref{tab:delA-tt}. \ First, the lepton cuts can
be loosened by observing leptons out to a pseudorapidity of 2.0
instead of 1.1, which is already possible for muons. This adds
$\sim65\%$ to the observed rate. Lowering the $p_{T}$ cut of only one
lepton to 10~GeV adds another $15\%$. Another few percent improvement
will automatically come from events where one or both $W$ bosons decay
to a tau lepton, which then decays to an electron or muon.  Imposing
looser lepton cuts but ignoring the contribution from taus, we obtain
a rate enhancement of about a factor of $1.9$ so that the uncertainty
decreases slightly, to $\delta
A_{FB}^{\ell}/A_{FB}^{\ell}\sim1.0$. The measurement becomes $1\
\sigma$ as indicated in the second line of Table
\ref{tab:delA-tt}.

We also expect that the $b$-tag efficiency should be improved
noticeably, both by adding kinematic track information in a neural net
analysis to the vertex tag, and by beginning to include soft lepton
tags, which is about $10\%$ per $b$~\cite{demina}. Together, these
highly likely improvements to $b$-jet tagging would enlarge the sample
by another factor of $1.9$. Third, Run~II could achieve its
\textquotedblleft stretch\textquotedblright\ goal of about
8~fb$^{-1}$, which would double the sample. Taken together (the most
optimistic scenario), these improvements would result in a dilepton
sample more about seven times what our conservative scenario
predicts. In this case, the asymmetry in the dilepton channel is still
challenging, yielding a $2.0\ \sigma$ measurement of the inclusive
asymmetry as indicated in the fourth line of Table \ref{tab:delA-tt}.

It is fair to ask how feasible these improvements are. The expanded
lepton pseudorapidity coverage is a safe bet, while the
lowered-$p_{T}$ cut is possible but not guaranteed (and also doesn't
result in much gain, compared to increased angular coverage). The
$b$-jet tagging efficiency will definitely improve noticeably with the
addition of soft lepton tags, while the prospects for improved silicon
vertex tags are not yet fully understood. We cannot guess at the total
integrated luminosity of Run~II, but anticipate continued efforts at
improvement.

Another natural possibility to consider is requiring only one $b$-tag,
which with the assumed improved tagging efficiency would increase the
sample by about a factor of $2.2$. While this approach introduces some
SM background, the background contribution will not be
large. Ultimately this must be taken into account, although it is
beyond the scope of this paper. Ignoring the small contribution from
background, a single $b$-tag strategy would result in an uncertainty
of $\delta A_{FB}^{\ell}/A_{FB}^{\ell}\sim0.34$ and yield a $2.9$
$\sigma$ measurement as indicated in the fifth line in Table
\ref{tab:delA-tt}. We note that it may also be possible to usefully study 
a zero $b$-tag dilepton sample. Half this sample, containing an $e+\mu$
mixed-flavor pair, has extremely low backgrounds. The primary
background for the other half, $ee$ or $\mu\mu$ events, arises from
the Drell-Yan process with an intermediate $Z$ boson with a
substantial ($\sim20\%$) asymmetry. We expect this background can be
controlled by cutting away the $Z$-pole region and demanding extra
jets.

We also mention one other possible scheme for improving the
measurement in the dilepton sample. Instead of an asymmetry based on a
single lepton being forward versus background in the laboratory frame,
consider one based on the longitudinal ordering of the 2 leptons,

\begin{equation}
{\bar{A}}_{FB}^{\ell}\,=\,\frac{N(\eta_{\ell^{+}}>\eta_{\ell^{-}
})-N\,(\eta_{\ell^{+}}<\eta_{\ell^{-}})}{N(\eta_{\ell^{+}}>\eta_{\ell^{-}
})+N\,(\eta_{\ell^{+}}<\eta_{\ell^{-}})}\;\,. \label{eq:nasym-ll}
\end{equation}
A preliminary look at the inclusive asymmetry using \textsc{mcfm}
\cite{MCFM} suggests that ${\bar{A}}_{FB}^{\ell}$ may be $50\%$
larger than $A_{FB}^{\ell}$. \ While the dilepton sample is likely not
as statistically significant as lepton+jets, this observable may
become more relevant if systematic uncertainties prove to be the
larger problem.

\begin{table}[th]
\begin{center}
\begin{tabular}
[c]{|c|c|c|c|c|c|c|c|}\hline
Sample & $\;\int\mathcal{L}dt$ [fb$^{-1}$]$\;$ & $\;\;$ cuts $\;\;$ &
$\;\;\;\epsilon_{2b}\;\;\;$ & $\;\;N^{\ell^{+}}_{F}+N^{\ell^{-}}_{B}\;\;$ &
$\;\;N^{\ell^{+}}_{B}+N^{\ell^{-}}_{F}\;\;$ & $\;\;\;\frac{\delta A_{FB}%
^{\ell}}{A_{FB}^{\ell}}\;\;\;$ & signif.\\\hline
dilepton & 4 & tight & 0.20 & 188 & 174 & 1.4 & 0.7 $\sigma$\\
dilepton & 4 & loose & 0.20 & 360 & 334 & 1.0 & 1.0 $\sigma$\\
dilepton & 4 & loose & 0.38 & 685 & 635 & 0.72 & 1.4 $\sigma$\\
dilepton & 8 & loose & 0.38 & 1370 & 1270 & 0.51 & 2.0 $\sigma$\\
dilepton & 8 & loose & 0.62$^{\ast}$ & 3074 & 2849 & 0.34 & 2.9 $\sigma
$\\\hline
lepton+jets & 4 & tight & 0.20 & 570 & 528 & 0.79 & 1.3 $\sigma$\\
lepton+jets & 4 & loose & 0.20 & 864 & 801 & 0.64 & 1.6 $\sigma$\\
lepton+jets & 4 & loose & 0.38 & 1642 & 1522 & 0.47 & 2.1 $\sigma$\\
lepton+jets & 8 & loose & 0.38 & 3285 & 3044 & 0.33 & 3.0 $\sigma$\\\hline
combined & 4 & tight & 0.20 & 757 & 702 & 0.69 & 1.5 $\sigma$\\
combined & 4 & loose & 0.20 & 1225 & 1135 & 0.54 & 1.8 $\sigma$\\
combined & 4 & loose & 0.38 & 2327 & 2157 & 0.39 & 2.5 $\sigma$\\
combined & 8 & loose & 0.38 & 4654 & 4314 & 0.28 & 3.6 $\sigma$\\
combined & 8 & loose & 0.38$^{\ast}$ & 6359 & 5893 & 0.24 & 4.2 $\sigma
$\\\hline
\end{tabular}
\vspace{-5mm}
\end{center}
\caption{Numbers of forward and backward lepton events, combining 
$\ell^{+}$ and $\ell^{-}$ samples (rounded to the nearest integer) and
expected statistical uncertainty on the measured lepton asymmetry,
$A^{\ell}_{FB}$, for inclusive $t\bar{t}$ production in $p\bar{p}$
collisions at Tevatron Run~II, $\sqrt{s}=2.0\,${TeV}, for top quark
mass $m_{t}=178\,${GeV, summed over the two detectors}. The lepton
asymmetry in each case is assumed to be that of the top quarks
themselves, $+3.8\%$. The upper block contains the result for the
dilepton sample, the middle block for the lepton+jets sample, and the
lower block for the combined samples. The total NLO cross section is
taken to be 7.4~pb, the average of the results of
Refs.~\cite{tt-NLO+}. \textquotedblleft Tight\textquotedblright\ cuts
refers to those of Eq.~\ref{eq:cuts}, while
\textquotedblleft loose\textquotedblright\ refers to the increased 
acceptance scenario described in
Secs.~\ref{subsec:incl-ll},\ref{subsec:incl-lh}. The $^{\ast}$ entries
represent using a single $b$-tag strategy for the dilepton sample.}
\label{tab:delA-tt}
\end{table}
%

%%%%%%%%%%%%%%%%%%%%%%%%%%%%%%%%%%%%%%%%%%%%%%%%%%%%%%%%%%%%%%%%%%%%%%%%

\newpage

\subsection{Lepton+jets final state \label{subsec:incl-lh}}

A different approach to increasing the sample size is to consider the
lepton+jets channel, which is slightly more than a factor six larger
in branching ratio than the dilepton channel. After cuts, the LO cross
section times NLO K-factor is 686~fb, approximately maintaining the
ratio of lepton+jets to dilepton rates. This channel does have a
larger background than dileptons, mostly from $W$+jets, which is not
fully understood either theoretically or from data. However, with
double $b$-jet tagging, S:B is about 6:1~\cite{SBratio}, so for our
purposes we assume a double $b$-tagging strategy and ignore background
contamination. Naturally, the inherent asymmetries of the same final
state from $W$+jets, about $20\%$ in the untagged sample, must
eventually be included, using exact matrix elements also for the
background.

Again conservatively assuming a double $b$-jet tagging efficiency of
$20\%$, we predict about 1100 events in 4~fb$^{-1}$ summed over both
experiments. Note that for this channel each event only contributes
once (a single lepton) to the measurement of the asymmetry, for a net
improvement over the dilepton case of about a factor of $3$. \ As
indicated in the middle block of Table~\ref{tab:delA-tt} this provides
for an initial measurement: $\delta
A_{FB}^{\ell}/A_{FB}^{\ell}\sim0.8$ and statistical significance of
$1.3$ $\sigma$. As above we consider the impact of increasing the
lepton acceptance range to $\eta(\ell)<2.0$, lowering the lepton
$p_{T}(j)$ cut to 15~GeV, and allowing for non-$b$ jets from $W$ decay
also down to $p_{T}>15$~GeV. \ This would increase the cross section
by about $50\%$ as indicated in the middle block of
Table~\ref{tab:delA-tt}. Prospects for improving this sample's size
from other sources, such as improved $b$-tagging and greater
integrated luminosity, are the same as in the dilepton sample. In the
most optimistic scenario of 8~fb$^{-1}$ per experiment, a double
$b$-tagging efficiency of $38\%$ and utilizing the looser cuts, we
find that the uncertainty in the asymmetry would be about 1/3, leading
to a $3\sigma$ measurement as indicated in
Table~\ref{tab:delA-tt}. Combining this and the dilepton channels
would increase the statistical significance to more than $4\sigma$ as
indicated in the bottom section of Table~\ref{tab:delA-tt}.

%%%%%%%%%%%%%%%%%%%%%%%%%%%%%%%%%%%%%%%%%%%%%%%%%%%%%%%%%%%%%%%%%%%%%%%%
%%%%%%%%%%%%%%%%%%%%%%%%%%%%%%%%%%%%%%%%%%%%%%%%%%%%%%%%%%%%%%%%%%%%%%%%

\section{Differential asymmetry in top quark pair{ }+1 jet events
\label{sec:tt1j}}

Next we consider correlations between the lepton asymmetry and a
single extra jet in the event not arising from the top quark
decays. As in the inclusive case, for all our calculations we use
matrix elements for $p\bar{p}\to t\bar{t}j$ production
($j=q,\bar{q},g$ summed over four light quark flavors), including
decay spin correlations and treating the intermediate $W$ bosons
off-shell, as generated by
\textsc{madgraph}~\cite{Stelzer:1994ta}. The extra parton does not
come from either of the on-shell top quark decays,
\textit{i.e.}, we do not include radiative decays. We calculate cross 
sections at $\sqrt{s}=2.0$~TeV requiring the additional hard jet to
have $p_{T}>20$~GeV and $|\eta|<2.0$, and use factorization and
renormalization scale choices of $\mu_{f}=\mu_{r}=m_{t}$, which yields
a perturbatively well-behaved result: before top quark decays, the
cross section is 1.03~pb, compared to 6.7-8.0~pb (avg. 7.4~pb) for the
NLO+(N)NLL $t\bar{t}$ rate~\cite{tt-NLO+}.

As noted in the previous section, choosing a more \textquotedblleft
physical\textquotedblright\ scale which takes into account the
softness of the additional parton, the cross section rises to 2.0~pb,
still a reasonable, perturbative result. The uncertainty on the
inclusive $t\bar{t}j$ rate awaits a full NLO
calculation~\cite{ttj-NLO}, but our conservative choice of $\mu=m_{t}$
gives the lowest cross section and may be regarded as a conservative
baseline. We will consider the possibility of larger normalization (by
a factor of 2) in our numerical results. Note that our predictions for
the asymmetry will not depend on the normalization, as discussed in
the next section.

%%%%%%%%%%%%%%%%%%%%%%%%%%%%%%%%%%%%%%%%%%%%%%%%%%%%%%%%%%%%%%%%%%%%%%%%

\subsection{Dilepton final state \label{subsec:1j-ll}}

We assume that both $b$ jets are tagged, so it is trivial to identify
the additional parton. The additional parton in the data, however,
will not always arise from the $t\bar{t}j$ \textquotedblleft
production-radiation\textquotedblright\ process as we have defined it.
Approximately $50\%$ of the time, \textit{i.e.}, at a rate
approximately equal to what we calculate here, the extra jet will be
due to radiative top quark decay~\cite{Orr:1994na}. In the dilepton
sample, this occurs when one of the $b$ quarks emits a hard
gluon. This defines the \textquotedblleft
decay-radiation\textquotedblright\ contribution. (As argued in
Ref.~\cite{Orr:1994na} this separation between production- and
decay-radiation, while approximate, is numerically reliable.) \ We
discuss later how the production and decay contributions might be
separated. As the calculations here describe the appearance of extra
radiation in the production mechanism, they will exhibit the asymmetry
characteristic of the LO $t\bar{t}j$ inclusive sample.

\begin{figure}[th]
\begin{center}
\includegraphics[scale=0.8]{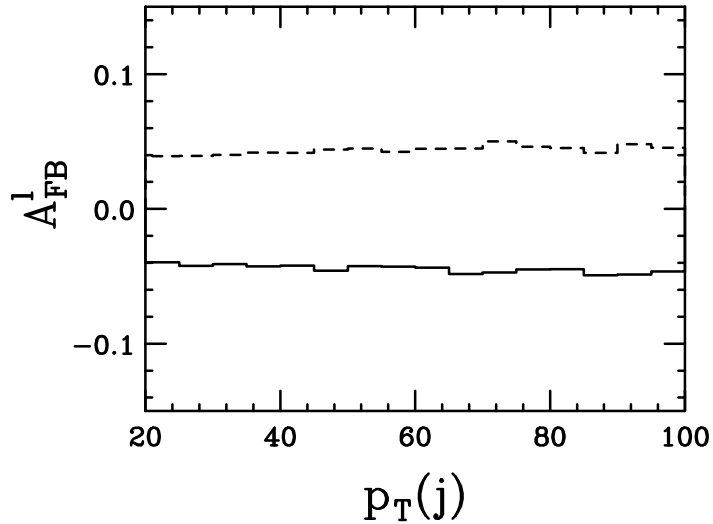} \vspace{-7mm}
\end{center}
\caption{Forward-backward lepton asymmetry in production-radiation 
$t\bar{t}j$ dilepton events as a function of the transverse momentum
of the additional hard jet. The $\ell^{+}$ ($\ell^{-}$) distribution
is the solid (dashed) curve. The two curves are $CP$-invariant up to
the level of Monte Carlo statistical uncertainty.}
\label{fig:asym-pT.tt1j.ll}
\end{figure}

We could again show the overall asymmetry, as in Sec.~\ref{sec:prod},
integrated over all kinematics of the additional jet, but this is just
a single number, with magnitude about $4-5\%$, the same as the
asymmetry of the top quarks themselves. We do note that when the top
quarks decay, leptons can be thrown preferentially back into the
hemisphere opposite to the top quark flight direction, due to the spin
of the top quark. Thus, the lepton asymmetry will not necessarily be
the same as the top quark asymmetry show in Sec.~\ref{sec:prod} before
decays. In any case, the more interesting result lies in the asymmetry
as a function of the kinematics of the additional jet.  We first show
$A_{FB}^{\ell}$ as a function of the $p_{T}$ of the extra jet in
Fig.~\ref{fig:asym-pT.tt1j.ll}. The corresponding version of
Eq.~\ref{eq:asym-ll}
\bq\label{asympt}
A_{FB}^{\ell}\,\left( p_{T}\left( j \right) \right) 
=\,\frac{N(\eta_{\ell}>0,p_{T}\left( j\right) )
-N\,(\eta_{\ell}<0,p_{T}\left( j\right) )}
{N(\eta_{\ell}>0,p_{T}\left( j\right) )
+N\,(\eta_{\ell}<0,p_{T}\left(j\right) )},
\eq
where, for example, $N(\eta_{\ell}>0,p_{T}\left( j\right) )$ is the
number of events with a forward lepton and an extra jet with
transverse momentum $p_{T}\left( j\right)$. As indicated in
Fig.~\ref{fig:asym-pT.tt1j.ll} this distribution is essentially
constant at the overall asymmetry value. Despite the lack of
structure, this result is important because it shows that the
asymmetry is not dependent on the extra parton $p_{T}$ cutoff, so our
results are independent of uncertainty due to the choice of the
perturbative cutoff in $p_{T}$. \ $CP$ still tells us that
$A_{FB}^{\ell^{+}}\left( p_{T}\left( j\right) \right)
=-A_{FB}^{\ell^{-}}\left( p_{T}\left( j\right) \right) $ as shown in
the figure. \ Note also that, as expected, $A_{FB}^{\ell^{+}}<0$, just
as for the LO $t\bar{t}j$ inclusive sample.

\begin{figure}[th]
\begin{center}
\includegraphics[scale=0.8]{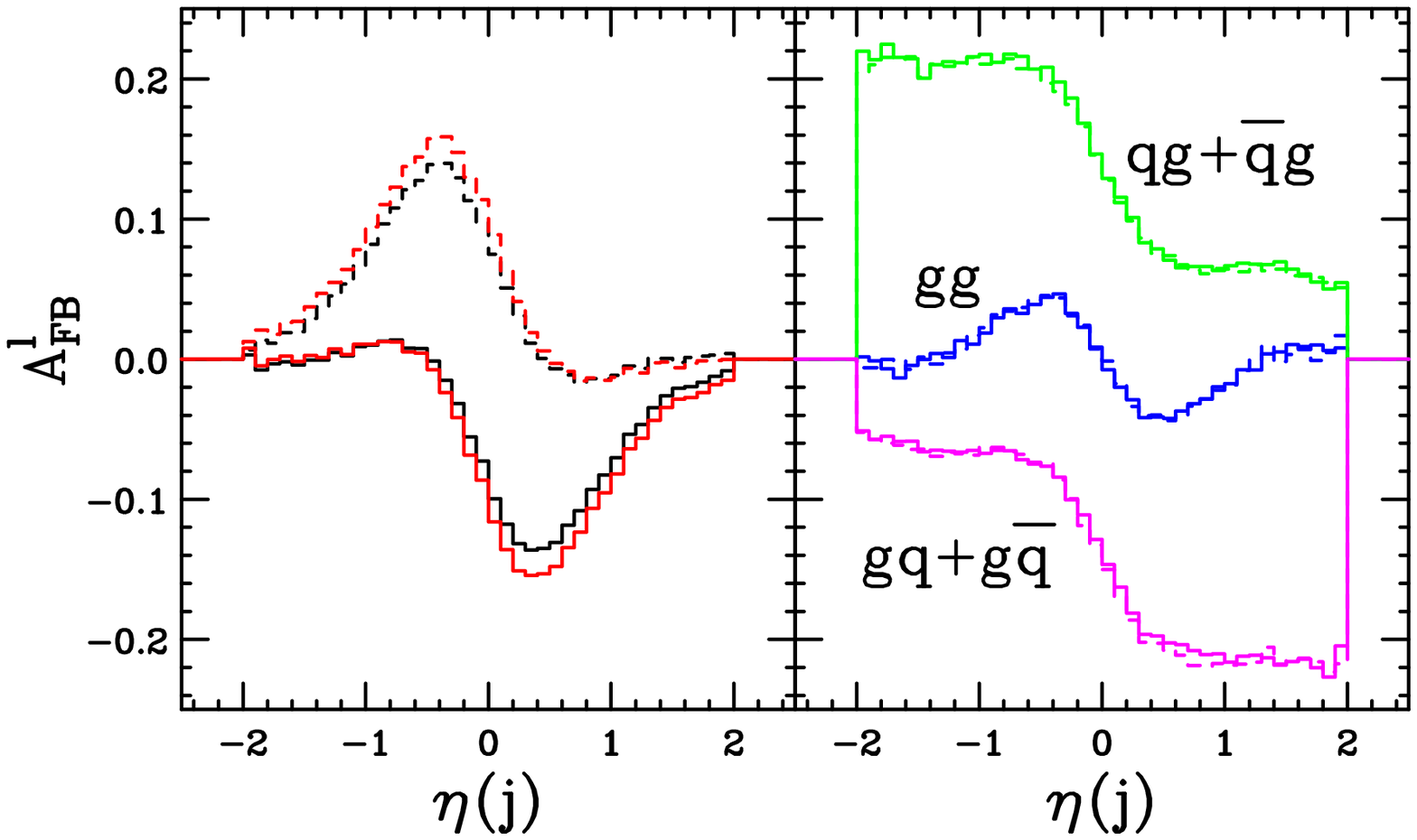} \vspace{-7mm}
\end{center}
\caption{Forward-backward lepton asymmetries for each subprocess in
production-radiation $t\bar{t}j$ dilepton events as a function of the
pseudorapidity of the additional hard jet. The $\ell^{+}$ ($\ell^{-}$)
distributions are shown by the solid (dashed) curves. The left panel
shows the dominant $q\bar{q}$ contribution to the asymmetry (the
curves with the slightly larger magnitude for both $\ell^{+}$ and
$\ell^{-}$) and the asymmetry for both charges for the total rate (the
curves with the smaller magnitude). \ The small difference between
$q\bar{q}$ and total arises from the contributions of the other parton
channels., whose asymmetries are indicated in the right panel. \ Note
the absence of any charge dependence in the curves in the right
panel.}
\label{fig:asym-y.tt1j.ll}
\end{figure}

The asymmetry with respect to the angular distribution of the
additional jet is far more revealing. \ The distribution of interest
is now
\bq
A_{FB}^{\ell}\,\left(  \eta\left(  j\right)  \right)  
=\,\frac{N(\eta_{\ell}>0,\eta\left(  j\right)  )
-N\,(\eta_{\ell}<0,\eta\left(  j\right)  )}
{N(\eta_{\ell}>0,\eta\left(  j\right)  )
+N\,(\eta_{\ell}<0,\eta\left(j\right)  )}, \label{asymeta}
\eq
where $CP$ invariance specifies that $A_{FB}^{\ell^{+}}\left(
\eta\left( j\right) \right) =-A_{FB}^{\ell^{-}}\left( -\eta\left(
j\right) \right) $, \textit{i.e.}, $A_{FB}^{\ell^{+}}$ and
$A_{FB}^{\ell^{-}}$ are mirror images of each other about
$\eta(j)=0$. \ We show this asymmetry distribution in
Fig.~\ref{fig:asym-y.tt1j.ll}, separately for the different parton
subprocess and for the combined result. \ Note that for each pair of
distributions (for the 2 lepton charges) the $CP$ relation just noted
is satisfied, but not always in the same way. \ For the subprocess
$gg\to t\bar{t}g$ the initial state is symmetric with respect to $C$
alone requiring that $A_{FB}^{\ell^{+}}\left( \eta\left( j\right)
,gg\right) =A_{FB} ^{\ell^{-}}\left( \eta\left( j\right) ,gg\right) $,
so that $CP$ now requires that the distributions are separately odd in
the jet rapidity, $A_{FB}^{\ell}\left( \eta\left( j\right) ,gg\right)
=-A_{FB}^{\ell}\left( -\eta\left( j\right) ,gg\right) $. \ These
constraints are in agreement with the corresponding (middle) curves in
the right panel of Fig.~\ref{fig:asym-y.tt1j.ll}. \ The specific form
of these distributions suggests that for this subprocess the
$t\bar{t}$ pair tends to recoil from the extra jet,\textit{ i.e.}, the
$t$ and $\bar{t}$ quarks (and the leptons from their decays) tend to
be in the same hemisphere while the recoiling jet is in the opposite
hemisphere (a backward jet yields a positive asymmetry and
conversely). \ Formally, the subprocesses containing a single quark
line exhibit an inclusive asymmetry, as found in Ref.~\cite{Kuhn},
although it is numerically extremely small. For the discussion here we
note that the combinations $qg+\bar{q}g$ or $gq+g\bar{q}$, as in the
upper and lower curves in the right panel of
Fig.~\ref{fig:asym-y.tt1j.ll},\ are strictly not $C$ symmetric\ states
due to the differences between $q$ and $\bar{q}$ distributions within
a proton (or antiproton). \ Thus it does
\emph{not} follow that $A_{FB}^{\ell^{+}}\left(  \eta\left(  j\right)
,qg+\bar{q}g\right)  =A_{FB}^{\ell^{-}}\left(  \eta\left(  j\right)
,qg+\bar{q}g\right)  $, $A_{FB}^{\ell^{+}}\left(  \eta\left(  j\right)
,gq+g\bar{q}\right)  =A_{FB}^{\ell^{-}}\left(  \eta\left(  j\right)
,gq+g\bar{q}\right)  $, even though these relations are approximately
numerically true in our results in Fig.~\ref{fig:asym-y.tt1j.ll}. \ So in this
case the (exact) constraint of $CP$ leads\ only to the approximate relation
$A_{FB}^{\ell}\left(  \eta\left(  j\right)  ,qg+\bar{q}g\right)
\simeq-A_{FB}^{\ell}\left(  -\eta\left(  j\right)  ,gq+g\bar{q}\right)
$. \ These features are illustrated in Fig.~\ref{fig:asym-y.tt1j.ll},
where the boost effects from the vastly different average Feynman $x$
values for quarks and gluons in the proton (antiproton) result in the
specific shapes of the curves. Qualitatively the $qg+\bar{q}g$
subprocess tends to a produce a $t\bar{t}$ pair in the forward
hemisphere ($A_{FB}^{\ell}>0$) with the extra jet recoiling into the
backward hemisphere. \ Forward/backward are reversed for the
$gq+g\bar{q}$ subprocess. \ Note that, after summing the $gg$, $qg$
and $gq$ subprocesses, their net asymmetry integrates to approximately
zero (over a symmetric interval in $\eta(j)$). \ Finally consider the
$q\bar{q}$ subprocess, which not only dominates the total rate at
Tevatron energies, but also exhibits the most interesting structure. \
The $q\bar{q}$ initial state is not separately invariant under $C$ and
$P$ so that we have $A_{FB}^{\ell^{+}}\left( \eta\left( j\right)
,q\bar{q}\right) \neq A_{FB}^{\ell^{-}}\left( \eta\left( j\right)
,q\bar{q}\right) $, while $CP$ still guarantees that
$A_{FB}^{\ell^{+}}\left( \eta\left( j\right) ,q\bar{q}\right)
=-A_{FB}^{\ell^{-}}\left( -\eta\left( j\right) ,q\bar{q}\right)$. \
The corresponding asymmetry distributions are illustrated in the left
panel of Fig.~\ref{fig:asym-y.tt1j.ll} for both the $q\bar{q}$
subprocess (the curves with large magnitudes) and the total of all
subprocesses (the curves with slightly smaller magnitudes). \ Note
that for these distributions the asymmetry for a single lepton charge
clearly does not integrate to zero. \ Qualitatively the $q\bar{q}$
subprocess tends to produce a $t$ quark in the backward hemisphere and
a $\bar{t}$ antiquark at rest in the lab when the extra jet is
forward, while a backward jet tends to correlate with a $\bar{t}$
antiquark in the forward hemisphere and a $t$ quark at rest in the
lab.

As indicated in Fig.~\ref{fig:asym-y.tt1j.ll} the asymmetry reaches a
maximum absolute value of around $14\%$ for extra jet rapidity in the
region around $|\eta(j)|\sim0.4$, and, even averaged over the region
$0<\eta(j)<1$ (for $\ell^{+}$), the asymmetry is over $10\%$. The
question of whether this can actually be measured is more difficult to
answer. The dilepton production-radiation $t\bar{t}j$ cross section at
LO with the \textquotedblleft tight\textquotedblright\ cuts of
Eq.~\ref{eq:cuts} is 15.1~fb. With 4~fb$^{-1}$ and a conservative
double $b$-tag rate of $20\%$, Run~II will then observe about $24$
dilepton production-radiation $t\bar{t}j$ double-$b$-tagged events for
D\O \ and CDF combined, based on the LO rate (and an approximately
equal number of radiative $t\bar{t}j$ events). This is not enough
statistics to perform even an overall asymmetry measurement, much less
in a restricted region of $\eta(j)$. \ However, with the prospects for
improving the sample size as discussed in Sec.~\ref{subsec:incl-ll},
as well as the uncertainty in cross section normalization, this
channel could possibly collect $\mathcal{O}$(100) events and would
then become interesting.

\begin{figure}[bh!]
\begin{center}
\includegraphics[scale=0.8]{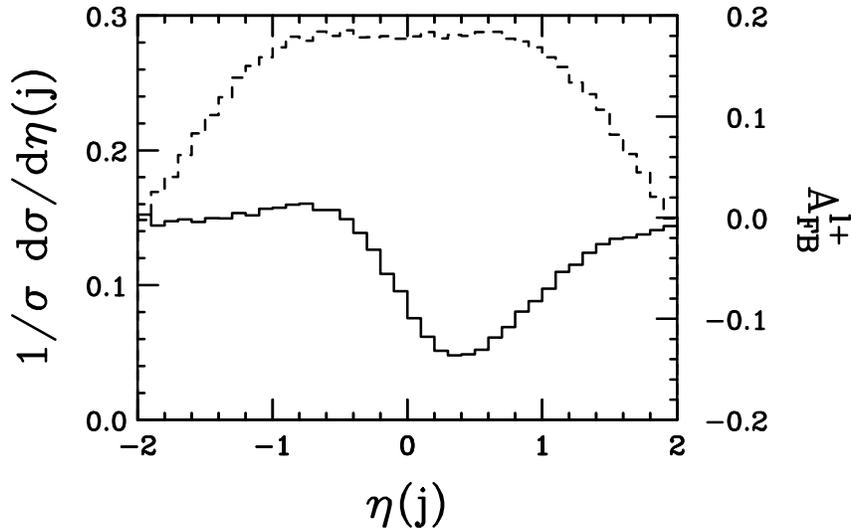} \vspace{-7mm}
\end{center}
\caption{Normalized differential cross section with respect to the extra 
jet pseudorapidity (dashed), overlaid with the total forward-backward
positive lepton asymmetry (solid), in dilepton production-radiation
$t\bar{t}j$ events with loose cuts as described in the text.}
\label{fig:overlay.1j.ll}
\end{figure}

We present Fig.~\ref{fig:overlay.1j.ll}, the normalized
angular-differential cross section distribution overlaid on the
(positive) lepton asymmetry, as a guide to how binning in $\eta(j)$
might be done depending on the ultimate statistics achieved in the
dilepton channel. Fairly central jet rapidity
features the largest asymmetry as well as large portion of the total
rate. At a minimum, one could use two bins, jet-forward and
jet-backward, discarding the region beyond $|\eta_{j}|\sim1$.

We have avoided the issue of the approximately equal-sized sample of
radiative decay $t\bar{t}j$ dilepton events. This is obviously a
non-trivial complication. Since the statistics of the dilepton channel
are weak to begin with, we don't attempt to address it further here,
except to note that radiative decay events are highly suppressed by
imposing an angular cut $\triangle R_{jb}\gg0.4$. For example, the cut
$\triangle R_{jb}>1.0$ reduces the radiative decay cross section by
more than a factor two~\cite{Orr:1994na}, but our calculations show
only a $20\%$ loss of production $t\bar{t}j$ dilepton events. In the
lucky circumstance that Run~II collects a very large dilepton top
quark sample, this issue should be thoroughly investigated.

%%%%%%%%%%%%%%%%%%%%%%%%%%%%%%%%%%%%%%%%%%%%%%%%%%%%%%%%%%%%%%%%%%%%%%%%

\subsection{Lepton+jets final state \label{subsec:1j-lh}}

We again also consider the lepton+jets channel, which is slightly more
than a factor six larger in branching ratio than the dilepton channel,
although it does suffer from a slight background as discussed in
Sec.~\ref{subsec:incl-lh}. After cuts, the LO cross section is
88.9~fb, slightly less than a factor six larger than the dilepton
channel with cuts.

The lepton+jets channel has one complication, that of correctly
identifying which of the three non-$b$-tagged jets is the
\textquotedblleft extra\textquotedblright\ one, \textit{i.e.,} the one
that is not a top quark decay product. This situation is aggravated by
\begin{figure}[bh!]
\begin{center}
\includegraphics[scale=0.8]{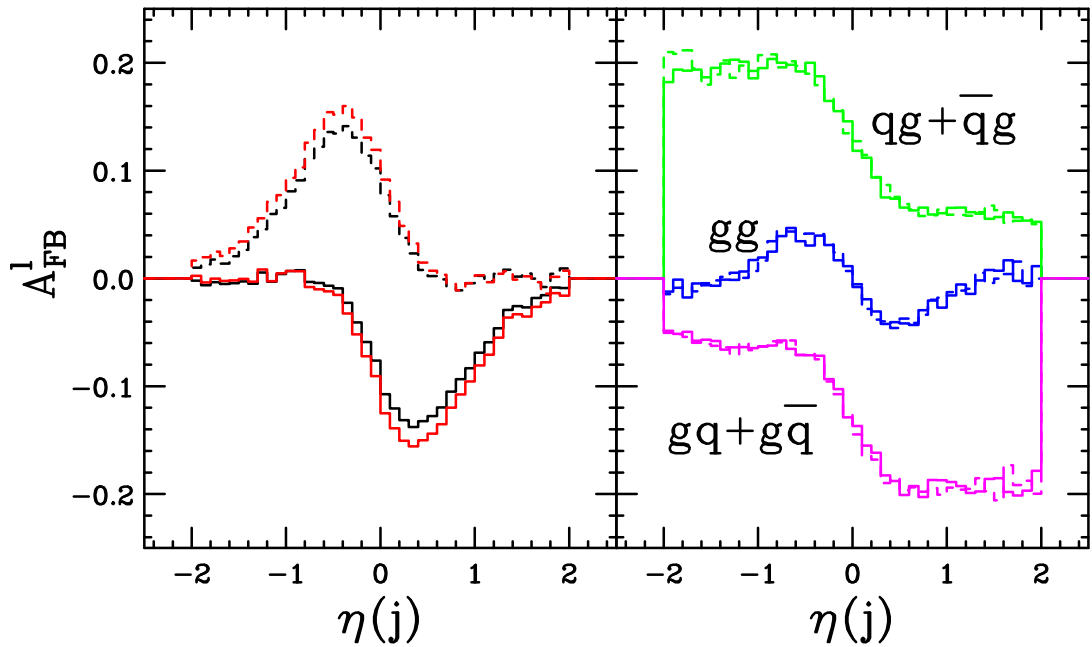} \vspace{-7mm}
\end{center}
\caption{Forward-backward lepton asymmetries for each subprocess in
production-radiation $t\bar{t}j$ lepton+jets events as a function of
the pseudorapidity of the additional hard jet. The $\ell^{+}$
($\ell^{-}$) distributions are shown by the solid (dashed) curves. The
left panel shows the dominant $q\bar{q}$ contribution to the asymmetry
(the curves with the slightly larger magnitude for both $\ell^{+}$ and
$\ell^{-}$) and the asymmetry for both charges for the total rate (the
curves with the smaller magnitude). \ The small difference between
$q\bar{q}$ and total arises from the contributions of the other parton
channels., whose asymmetries are indicated in the right panel. \ Note
the absence of any charge dependence in the curves in the right
panel.}
\label{fig:asym-y.tt1j.lh}
\end{figure}
the combinatorics, but can be largely addressed by imposing the $W$
mass constraint on one jet pair, the top quark mass constraint on that
pair and a $b$ jet, and the transverse top mass constraint on the
other $b$ jet, the lepton and the missing transverse energy.  We do
not examine this issue further here, but expect misidentification
effects due to kinematical combinatorics to be a minor
correction. This issue should of course be examined in detail when the
channel is studied with full detector simulation. Such mass
constraints also remove the vast bulk of top quark decay-radiation
events from the sample, as our minimum $p_{T}(j)$ requirement means
that sample contamination from radiative events enters mostly via jet
mismeasurement at the edges of invariant mass windows.

Fig.~\ref{fig:asym-y.tt1j.lh} shows the lepton asymmetry in
production-radiation $t\bar{t}j$ lepton+jets events, similar to
Fig.~\ref{fig:asym-y.tt1j.ll} for the dilepton case, also with the
cuts of Eq.~\ref{eq:cuts}. This plot assumes that the additional jet,
which did not come from a top decay, is determined with $100\%$
accuracy, as discussed above. The asymmetry features in this sample
are almost identical to those in the dilepton sample. The asymmetry as
a function of extra jet $p_{T}$ is flat as in the dilepton channel,
and we do not show the corresponding figure here.

Again starting from the assumption of $20\%$ efficiency for double
$b$-jet tagging (no improvement over the current level), our LO
estimate predicts about 142 events in 4~fb$^{-1}$ for both
experiments. Measuring the differential asymmetry therefore appears
possible only for large pseudorapidity bins. Applying the loose cuts
as in the inclusive sample gives about a factor 1.5 increase. Note
that we keep the requirement that the extra jet has $p_{T}>20$~GeV,
simply to keep our calculation confidently in the perturbative
regime. Prospects for improving this sample's size from other sources,
such as improved $b$-tagging, greater integrated luminosity, or cross
section normalization, are the same as in the dilepton sample.

%%%%%%%%%%%%%%%%%%%%%%%%%%%%%%%%%%%%%%%%%%%%%%%%%%%%%%%%%%%%%%%%%%%%%%%%

\vspace{-2mm}
\subsection{Numerical results \label{subsec:tt1j_sum}}

Here we make some simple estimates of the uncertainty that might be
achieved on an $A_{FB}^{\ell}$ measurement in the inclusive
$t\bar{t}j$ sample. The ideal objective in this case would be to map
out the asymmetry with respect to the angular structure of the
\begin{figure}[bh!]
\begin{center}
\includegraphics[scale=0.8]{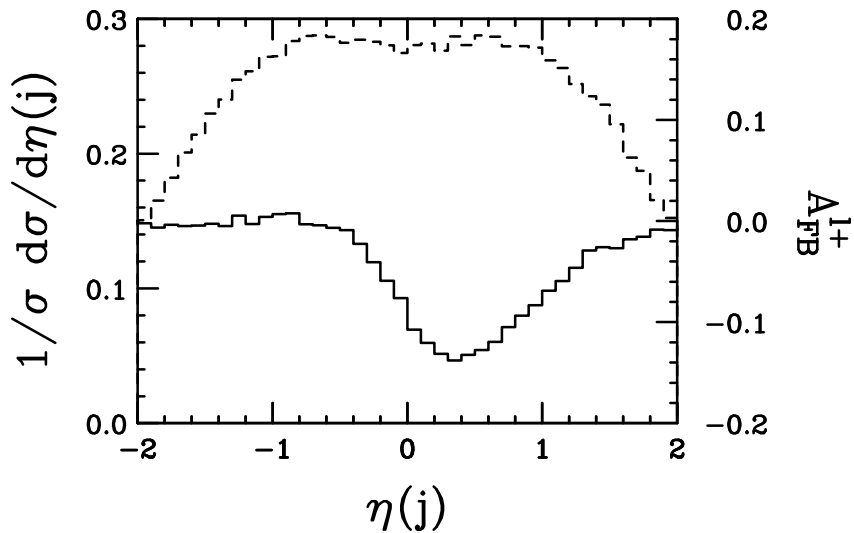} \vspace{-8mm}
\end{center}
\caption{Normalized differential $t\bar{t}j$ cross section with respect 
to the extra jet pseudorapidity (dashed), overlaid with the total
forward-backward positive lepton asymmetry (solid), in lepton+jets
events with loose cuts as described in the text.}
\label{fig:overlay.1j.lh}
\end{figure}
additional hard parton, which is additional information that the
inclusive and exclusive $t\bar{t}$ samples do not have.  We consider
both the dilepton and lepton+jets samples, although the latter has far
better statistics and is expected to largely avoid the complications
of radiative top quark decay. We consider only the statistical
uncertainty, as the most important systematic uncertainties are
related to detector effects, which we cannot include; or the very
small backgrounds in double $b$-tagged events, which are expected to
be a much smaller uncertainty than the limited statistics will allow
for.

Due to the limited statistics available, we select a single bin in
extra jet pseudorapidity of $-0.1<\eta(j)<1.1$ ($-1.1<\eta(j)<0.1$)
for $\ell^{+}$ ($\ell^{-}$). Varying this over a range of a few tenths
either direction at either limit does very little to the overall
uncertainty, while increasing the range to increase statistics results
in an overall lower asymmetry, and vice versa, as seen in
Fig.~\ref{fig:overlay.1j.lh}. The asymmetry in this bin is $11\%$.

We show our numerical results in Table~\ref{tab:delA-ttj}, which has a
general format similar to Table \ref{tab:delA-tt}. As suggested by
Figs.~\ref{fig:overlay.1j.ll} and~\ref{fig:overlay.1j.lh}
approximately $1/3$ of the events fall into the rapidity window noted
above, $-0.1<\eta(j)<1.1$ ($-1.1<\eta(j)<0.1$) for
$\ell^{+}$($\ell^{-}$). \ Thus, since the dilepton events still
contribute twice, the 24 dilepton events mentioned above yield $16$
counted leptons in the first line of the dilepton block ($24\times
1/3\times2$) in Table~\ref{tab:delA-ttj}.  For the lepton+jets sample
the number of leptons is just $1/3$ of the
\begin{table}[bh!]
\begin{center}
\begin{tabular}
[c]{|c|c|c|c|c|c|c|c|c|c|}\hline
Sample & $\;\int\mathcal{L}dt$ [fb$^{-1}$]$\;$ & $\;$ cuts $\;$ &
$\;\;\epsilon_{2b}\;\;$ & $\;A_{FB}^{\ell}\;$ & $\;N_{F}^{\ell^{+}}$
+$\;N_{B}^{\ell^{-}}\;$ & $\;N_{B}^{\ell^{+}}$+$\;N_{F}^{\ell^{-}}\;$ &
$\;K_{NLO}^{t\bar{t}j}\;$ & $\;\;\frac{\delta A_{FB}^{\ell}}{A_{FB}^{\ell}%
}\;\;$ & signif.\\\hline
dilepton & 4 & tight & 0.20 & -0.11 & 7 & 9 & 1 & 2.3 & 0.4 $\sigma$\\
dilepton & 4 & loose & 0.20 & -0.12 & 14 & 17 & 1 & 1.5 & 0.7 $\sigma$\\
dilepton & 4 & loose & 0.38 & -0.12 & 26 & 33 & 1 & 1.1 & 0.9 $\sigma$\\
dilepton & 8 & loose & 0.38 & -0.12 & 51 & 66 & 1 & 0.75 & 1.3 $\sigma$\\
dilepton & 8 & loose & 0.38 & -0.12 & 103 & 132 & 2 & 0.53 & 1.9 $\sigma
$\\\hline
lepton+jets & 4 & tight & 0.20 & -0.11 & 21 & 26 & 1 & 0.8 & 0.8 $\sigma$\\
lepton+jets & 4 & loose & 0.20 & -0.12 & 31 & 40 & 1 & 1.0 & 1.0 $\sigma$\\
lepton+jets & 4 & loose & 0.38 & -0.12 & 60 & 76 & 1 & 0.69 & 1.4 $\sigma$\\
lepton+jets & 8 & loose & 0.38 & -0.12 & 119 & 153 & 1 & 0.49 & 2.0 $\sigma$\\
lepton+jets & 8 & loose & 0.38 & -0.12 & 239 & 306 & 2 & 0.35 & 2.9 $\sigma
$\\\hline
combined & 4 & tight & 0.20 & -0.11 & 28 & 36 & 1 & 1.1 & 0.9 $\sigma$\\
combined & 4 & loose & 0.20 & -0.12 & 45 & 58 & 1 & 0.8 & 1.3 $\sigma$\\
combined & 4 & loose & 0.38 & -0.12 & 85 & 109 & 1 & 0.58 & 1.7 $\sigma$\\
combined & 8 & loose & 0.38 & -0.12 & 171 & 219 & 1 & 0.41 & 2.4 $\sigma$\\
combined & 8 & loose & 0.38 & -0.12 & 342 & 437 & 2 & 0.29 & 3.4 $\sigma
$\\\hline
\end{tabular}
\vspace{-5mm}
\end{center}
\caption{Numbers of forward and backward lepton events combining 
$\ell^{+}$ and $\ell^{-}$ samples (rounded to the nearest integer),
and expected statistical uncertainty on the absolute value of the
measured lepton asymmetry, $A_{FB}^{\ell^{+}}$, for
production-radiation $t\bar{t}j$ events in $p\bar{p}$ collisions at
Tevatron Run~II, $\sqrt{s}=2.0\,${TeV}, for top quark mass $m_{t}=178\
${GeV,} summed over the two detectors with $-0.1<\eta(j)<1.1$ for the
extra jet. $K_{NLO}^{t\bar{t}j}=\frac{\sigma_{NLO}}{\sigma_{LO}}$ for
the $t\bar{t}j$ rate. The upper block contains results for the
dilepton sample, the middle block for lepton+jets sample, and the
lower block for the combined samples. \textquotedblleft
Tight\textquotedblright\ cuts refers to those of Eq.~\ref{eq:cuts},
while \textquotedblleft loose\textquotedblright \ refers to the
possible increased acceptance as described in
Sec.~\ref{sec:recalc}. The first four lines in each block represent
improvements in detector and machine performance, while the last line
represents uncertainty in the $t\bar{t}j$ rate as discussed in
Sec.~\ref{sec:tt1j}.}
\label{tab:delA-ttj}
\end{table}
number of events so that $142$ events become $47$ leptons.  We have
also included a column for the value of the asymmetry,
$A_{FB}^{\ell^{+}}$, which is calculated directly from the cross
sections and not the rounded numbers of leptons in the Table, and a
column for the assumed $K$ factor. \ For our very conservative
baseline scenario with $\sigma_{NLO}/\sigma_{LO}=K=1$, the measurement
is effectively consistent with zero. The very likely scenario of
increased lepton acceptance and expected $b$-tagging improvement, line
3, allows for a more interesting measurement.  However, either
increased luminosity or a fortuitous larger cross section, illustrated
here with $\sigma_{NLO}/\sigma_{LO}=2$ corresponding to a smaller
choice of factorization/renormalization scale, would be required to
allow the statistical significance of the measurement to reach the
level of $3-4$ $\sigma$.

%%%%%%%%%%%%%%%%%%%%%%%%%%%%%%%%%%%%%%%%%%%%%%%%%%%%%%%%%%%%%%%%%%%%%%%%

\section{Asymmetry in top quark pair{ }+ 0 jet exclusive events
\label{sec:excl0j}}

We have seen that measuring the asymmetry in the inclusive $t\bar{t}$
sample may be doable but challenging, and that in the $t\bar{t}j$
inclusive production-radiation sample it will be extremely
difficult. The inclusive $t\bar{t}$ sample suffers from a smaller
asymmetry, while for the inclusive $t\bar{t}j$ sample statistics will
be a larger problem. As seen in Sec.~\ref{sec:prod}, events without an
additional hard jet should have a similar asymmetry to $t\bar{t}j$
events, although of opposite sign. Yet this sample would be
approximately $6$ times the size of the inclusive $t\bar{t}j$ sample,
before any cuts are put on the jet rapidity. An obvious strategy is to
experimentally attempt a jet veto on the $t\bar{t}$ inclusive sample
to obtain an exclusive $t\bar{t}0j$ sample with better statistics than
the inclusive $t\bar{t}j$ sample.

While definitely worth pursuing, this will not be
straightforward. Recall that approximately as many events in inclusive
$t\bar{t}$ production will have an additional hard jet from radiative
top quark decay as from production radiation. One does not want to
veto events with these jets. Certainly some fraction of the time D\O\
and CDF will be able to tell that the extra jet likely comes from
radiative decay, based on the invariant mass of two or three jets
reconstructing to a $W$ boson, or for a $b$ jet and the extra jet plus
either two jets or a lepton to have an invariant mass (or transverse
mass) equal to $m_{t}$. Determining how efficiently this can be done
is beyond the scope of this paper, so we will ignore the complication
of radiative top quark decays. Our goal in making estimates of how
well one might measure $A_{FB}^{\ell}$ in the exclusive sample is
simply to highlight the beneficial features of the larger asymmetry
and the greater statistics.

\begin{table}[th]
\begin{center}
\begin{tabular}
[c]{|c|c|c|c|c|c|c|c|c|c|}\hline
Sample & $\;\int\mathcal{L}dt$ [fb$^{-1}$]$\;$ & $\;$ cuts $\;$ &
$\;\;\epsilon_{2b}\;\;$ & $\;A_{FB}^{\ell}\;$ & $\;N_{F}^{\ell^{+}}$
+$\;N_{B}^{\ell^{-}}\;$ & $\;N_{B}^{\ell^{+}}$+$\;N_{F}^{\ell^{-}}\;$ &
$K_{NLO}^{t\bar{t}j}$ & $\;\;\frac{\delta A_{FB}^{\ell}}{A_{FB}^{\ell}}\;\;$ &
signif.\\\hline
dilepton & 4 & tight & 0.20 & 0.044 & 176 & 161 & 1 & 1.4 & 0.7 $\sigma$\\
dilepton & 4 & loose & 0.20 & 0.045 & 339 & 310 & 1 & 1.0 & 1.0 $\sigma$\\
dilepton & 4 & loose & 0.38 & 0.045 & 644 & 589 & 1 & 0.75 & 1.3 $\sigma$\\
dilepton & 8 & loose & 0.38 & 0.045 & 1288 & 1178 & 1 & 0.53 & 1.9 $\sigma$\\
dilepton & 8 & loose & 0.62$^{\ast}$ & 0.048 & 2891 & 2644 & 1 & 0.35 & 2.8
$\sigma$\\
dilepton & 8 & loose & 0.38 & 0.052 & 1207 & 1087 & 2 & 0.53 & 1.9 $\sigma
$\\\hline
lepton+jets & 4 & tight & 0.20 & 0.050 & 502 & 454 & 1 & 0.85 & 1.2 $\sigma$\\
lepton+jets & 4 & loose & 0.20 & 0.052 & 764 & 689 & 1 & 0.69 & 1.4 $\sigma$\\
lepton+jets & 4 & loose & 0.38 & 0.052 & 1452 & 1308 & 1 & 0.50 & 2.0 $\sigma
$\\
lepton+jets & 8 & loose & 0.38 & 0.052 & 2903 & 2616 & 1 & 0.35 & 2.8 $\sigma
$\\
lepton+jets & 8 & loose & 0.38 & 0.071 & 2521 & 2188 & 2 & 0.38 & 2.6 $\sigma
$\\\hline
combined & 4 & tight & 0.20 & 0.049 & 678 & 615 & 1 & 0.73 & 1.4 $\sigma$\\
combined & 4 & loose & 0.20 & 0.050 & 1103 & 999 & 1 & 0.57 & 1.7 $\sigma$\\
combined & 4 & loose & 0.38 & 0.050 & 2096 & 1897 & 1 & 0.42 & 2.4 $\sigma$\\
combined & 8 & loose & 0.38 & 0.050 & 4191 & 3794 & 1 & 0.29 & 3.4 $\sigma$\\
combined & 8 & loose & 0.38$^{\ast}$ & 0.048 & 5794 & 5260 & 1 & 0.25 & 4.0
$\sigma$\\
combined & 8 & loose & 0.38 & 0.065 & 3728 & 3275 & 2 & 0.32 & 3.2 $\sigma
$\\\hline
\end{tabular}
\vspace{-5mm}
\end{center}
\caption{Numbers of forward and backward lepton events combining $\ell^{+}$
and $\ell^{-}$ samples (rounded to the nearest integer), and expected
statistical uncertainty on the absolute value of the measured lepton
asymmetry, $A_{FB}^{\ell}$, for production-radiation-vetoed
$t\bar{t}0j$ exclusive events in $p\bar{p}$ collisions at Tevatron
Run~II, $\sqrt{s} =2.0\,${TeV}, for top quark mass $m_{t}=178\,${GeV,}
summed over the two
detectors. $K_{NLO}^{t\bar{t}j}=\frac{\sigma_{NLO}}{\sigma_{LO}}$ for
the $t\bar{t}j$ rate. The upper block contains results for the
dilepton sample, the middle block for the lepton+jets sample, and the
lower block for the combined samples. \textquotedblleft
Tight\textquotedblright\ cuts refers to those of Eq.~\ref{eq:cuts},
while \textquotedblleft loose\textquotedblright
\ refers to the possible increased acceptance as described in
Sec.~\ref{sec:recalc}. The first four lines in each block represent
improvements in detector and machine performance, while the last two
line represents uncertainty in the $t\bar{t}j$ rate as discussed in
Sec.~\ref{sec:tt1j}. The fifth line in the first and last blocks (the
$^{\ast }$ entries) represent using a single $b$-tag strategy for the
dilepton sample as discussed in Sec.~\ref{subsec:incl-ll}.}
\label{tab:delA-excl0j}
\end{table}

Our procedure is to take the numbers of events expected in the
inclusive $t\bar{t}$ sample based on the fraction of events that pass
the cuts, as in Sec.~\ref{sec:prod}, and subtract from those forward
and backward event numbers those from the $t\bar{t}j$ sample. The
obvious deficiency in this is that the inclusive asymmetry is known
only from a NLO calculation (necessarily) which does not include
decays, so the asymmetry shift from kinematical cuts effects cannot
yet be taken into account. Thus, one should not put too much faith in
the exact values predicted, as we expect future calculations to shift
them noticeably. Our calculations, however, do represent the state of
the art at this time.

We show our numerical results in Table~\ref{tab:delA-excl0j} where the
number of contributing events is approximately $18$ times larger than
in Table~\ref{tab:delA-ttj} due to the larger cross section
($\times6$) and the absence of a jet rapidity cut ($\times3$). \ As
before, we consider tight and loose levels of kinematic cuts, low and
high Tevatron integrated luminosity, pessimistic and optimistic
$b$-tagging efficiencies, and a possible 1-$b$-tag strategy for the
dilepton sample only. Note that if the $t\bar{t}j$ normalization is at
the upper end of our cross section estimate, a factor of two larger
than the conservative scale choice for the calculation, then the
statistical significance of the exclusive $t\bar{t}0j$ sample goes
\textit{down} slightly, due to the decreased sample size, which is partially
compensated by the larger remaining asymmetry. \ As expected, the
asymmetry in this sample is positive and somewhat larger in magnitude
than the fully inclusive expectation of $3.8\%$. \ While the
statistics are better than indicated in Table~\ref{tab:delA-ttj},
these results still suggest that with the standard luminosity, the
conservative cross section and loose cuts, only a $2.4$ $\sigma$
measurement is possible. \ An integrated luminosity greater than 4
fb$^{-1}$ will be required for a more accurate determination of the
asymmetry.

%%%%%%%%%%%%%%%%%%%%%%%%%%%%%%%%%%%%%%%%%%%%%%%%%%%%%%%%%%%%%%%%%%%%%%%%

\section{Differential asymmetry in top quark pair{ }+ 2 jet events
\label{sec:tt2j}}

The LO cross section for $t\bar{t}jj$ production at Tevatron Run~II
varies from 160 to 300~fb over the range of scale choices discussed
earlier. We again calculate this with exact matrix elements from
\textsc{madgraph}, choosing $p_{T}(j)>20$~GeV and $|\eta(j)|<2.0$ as
the limits in phase space integration for the extra jets. We further
\begin{figure}[th!]
\begin{center}
\includegraphics[scale=0.8]{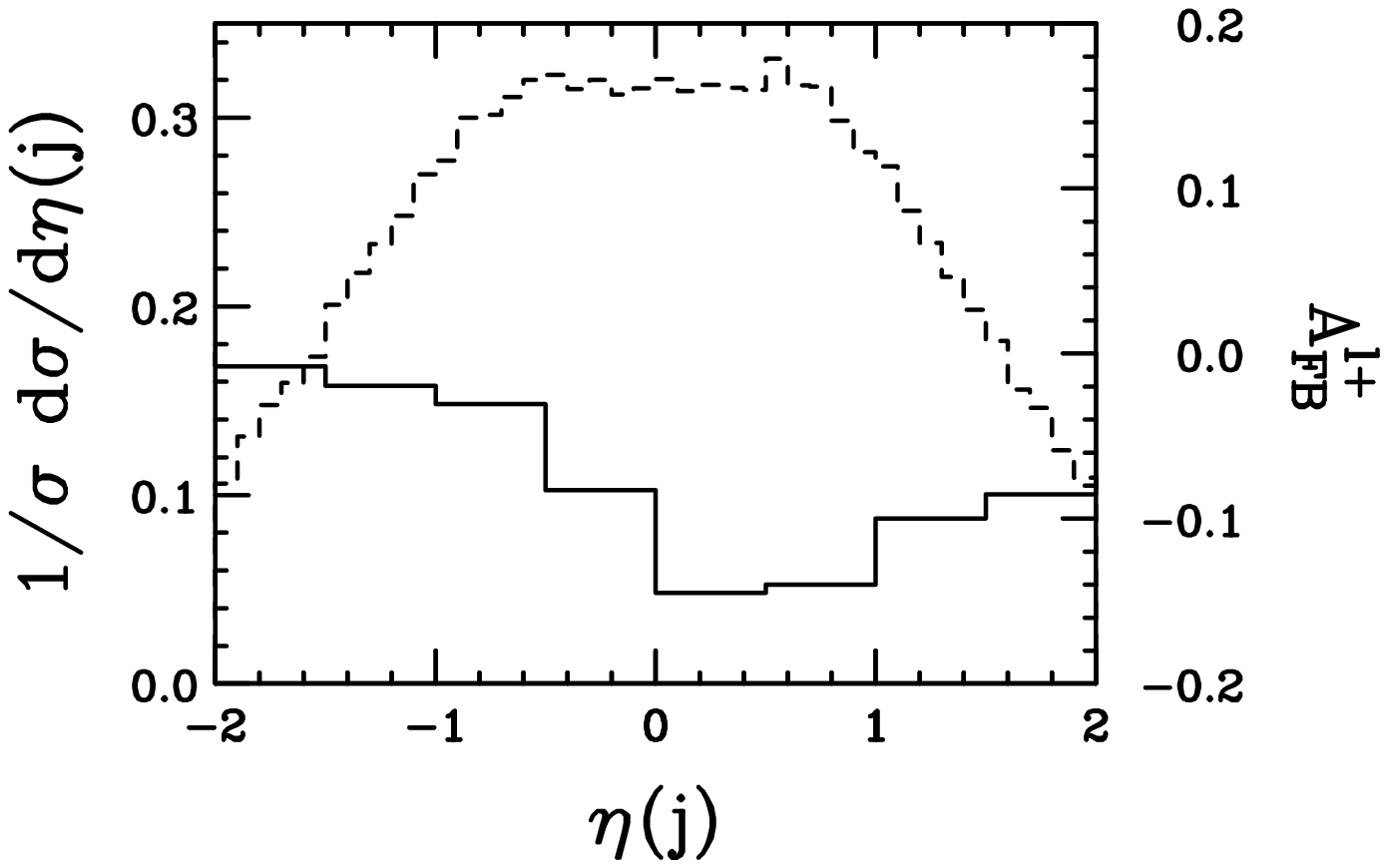} \vspace{-7mm}
\end{center}
\caption{Normalized differential $t\bar{t}jj$ cross section with respect 
to the higher-$p_{T}$ extra jet pseudorapidity (dashed), overlaid with
the total forward-backward positive lepton asymmetry (solid), in
lepton+jets events with loose cuts as described in the text.}
\label{fig:overlay.2j.lh}
\end{figure}
require $\triangle R_{jj}>0.5$ for the extra jet pair to be separated
from each other, to avoid the collinear singularity from gluon
splitting. Compared to the 1--2~pb cross section for $t\bar{t}j$
production, this is approximately a factor 5 lower in
statistics. Because of this, here we consider only the lepton+jets
channel with the looser cuts and the improved b-tagging scenario.

Similar to the plot in Sec.~\ref{subsec:1j-lh} we show the normalized
differential cross section with respect to the rapidity of the
higher-$p_{T}$ of the two extra jets, as well as the differential
asymmetry, in Fig.~\ref{fig:overlay.2j.lh}. The asymmetry shows
similar structure to that of Fig.~\ref{fig:overlay.1j.lh}, although
here we use larger binning as it takes considerably more computing
time to obtain the same level of statistics for smaller bins. The
optimal region to use to obtain the best statistical significance for
an asymmetry measurement is about $-0.5<|\eta(j)|<2.0$, which includes
approximately $2/3$ of the total rate. For 8~fb$^{-1}$ per experiment
and two combined experiments, with conservative cross section
normalization and a double b-tagging efficiency of $38\%$ Run~II
(summing the 2 detectors) could expect about 76 events with average
asymmetry $A_{FB}^{\ell^{+}}=-0.11$ over this region. This would yield
only about a $1.0\ \sigma$ measurement. If the cross section is really
a factor of two larger ($K=2$), this would still only be about a
$1.4\sigma$ measurement: interesting and worth pursuing, but difficult
to achieve a useful level of precision.

%%%%%%%%%%%%%%%%%%%%%%%%%%%%%%%%%%%%%%%%%%%%%%%%%%%%%%%%%%%%%%%%%%%%%%%%

\section{Comparison with Parton Shower Monte Carlo \label{sec:PSMC}}

Since Parton-shower Monte Carlo simulation packages, such as
\textsc{pythia}~\cite{pythia}, \textsc{herwig}~\cite{herwig} and 
\textsc{sherpa}~\cite{sherpa}, use only LO matrix elements, which do 
not exhibit the interference structure of the NLO matrix elements,
they are not expected to reproduce the
\textit{inclusive} asymmetry seen at NLO. There is, however, the 
possibility that certain choices of the parameters that control the
simulation of \textquotedblleft color-coherence\textquotedblright\ in
the showering/hadronization components of the Monte Carlo can lead to
correlations that mimic the $t\bar{t}j$ and $t\bar{t}0j$ asymmetries
observed in the NLO perturbative results. Recall that, while the
individual partons describing the short distance scattering carry
color charges, the long distance initial- and final-state hadrons are
all color neutral. \ The requirement of reassembling at long distances
into color-neutral states requires correlations between the partons.
Complete fixed-order perturbative calculations account explicitly for
the color-coherence, at fixed order, but Monte Carlo simulations
include color-flow correlations only approximately, based on various
models. \ To illustrate this, we have studied $t\bar{t}$ dilepton
production using \textsc{pythia}~\cite{pythia}.

We imposed the cuts of Eq.~\ref{eq:cuts} on the $b$ partons, leptons,
and missing transverse momentum in Pythia-generated $t\bar{t}$
dilepton events. To identify extra radiation, we used the
\texttt{PYCELL} subroutine with a cone size of $R=0.4$. Any
reconstructed jets were required to be $\Delta R>0.4$ away from the
$b$ partons to be considered \textquotedblleft extra
radiation\textquotedblright. \ As the matrix elements used by
\textsc{pythia} are LO, the lepton and antilepton distributions are
the same, and the fully inclusive asymmetry is zero. The situation is
more interesting when extra radiation is demanded or vetoed. In
particular, correlations can be introduced by the angular structure of
the all-orders QCD showering simulated in the Monte Carlo. \ For
example, in \textsc{pythia} the parameters
\texttt{MSTJ(50)} and \texttt{MSTP(67}) control the structure of the 
final state (\texttt{MSTJ(50)}) and initial state (\texttt{MSTP(67)})
showers. \ With the default values (the \textquotedblleft
on\textquotedblright\ value) for these two parameters, \textsc{pythia}
adds initial- and final-state radiation to events based on color-flow
information, while for the \textquotedblleft off\textquotedblright\
value the showers develop independently of the color information and
largely independently of the rest of the event. The color-flow
constraint for the \textquotedblleft on\textquotedblright\ case has
the effect of restricting radiation to appear preferentially in the
angular region between the $p$ direction and the produced top quark
direction, or between the $\bar{p}$ direction and the anti-top quark
direction. \ Thus the presence of extra radiation tends to be
correlated with larger angle scattering of the top quark,
\textit{i.e}., the top quark tends to be in the backward hemisphere
when radiation is present. \ This correlation is intended to
approximate the structure\ expected as a result of the
\textquotedblleft color coherence\textquotedblright\ in a full, all
orders matrix element analysis with color singlet asymptotic states. 
\ Not only can this feature of \textsc{pythia} lead to a non-zero lepton
forward-backward asymmetry in the presence of extra radiation (even
when integrated over the rapidity of the radiation), but this
asymmetry has the same sign and general correlation with the radiation
as in the earlier $t\bar{t}j$ perturbative analysis, as we now
explore. Note that, by implication, the complementary event sample
(with a veto on radiation) has an asymmetry of the opposite sign if
the entire sample is to have a zero inclusive asymmetry. This means
that the exclusive sample in
\textsc{pythia} will have an asymmetry of the same sign as the
$t\bar{t}0j$ exclusive sample.

\begin{figure}[th!]
\begin{center}
\includegraphics[scale=0.83]{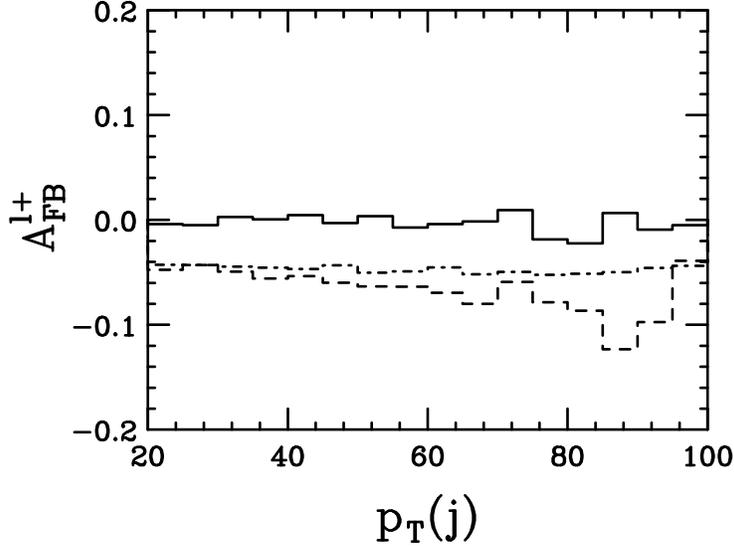} \vspace{-7mm}
\end{center}
\caption{Asymmetry $A_{FB}^{\ell^{+}}$ as a function of extra jet 
$p_{T}$ with \textsc{pythia} color-flow correlations turned off
(solid) and on (dashed), as compared with the matrix element
prediction (dot-dashed).}
\label{fig:ptasym_pythia2}
\end{figure}

As in our earlier discussion of the matrix element calculations, $CP$
invariance requires that the $\ell^{-}$ distribution is always
mirror-antisymmetric to that of the $\ell^{+}$ about $\eta(j)=0$,
$A_{FB}^{\ell^{+}}\left( \eta\left( j\right) \right)
=-A_{FB}^{\ell^{-}}\left( -\eta\left( j\right) \right)$. \ If the
color-flow correlations in \textsc{pythia} are turned
\textquotedblleft off\textquotedblright \ (\texttt{MSTP(67)} and 
\texttt{MSTJ(50)} \textquotedblleft off\textquotedblright), the final 
state has no knowledge of the $C$ (non-invariant) structure of the
initial state and (as in the earlier discussion)
$A_{FB}^{\ell^{+}}\left( \eta\left( j\right) ,\text{off} \right)
=A_{FB}^{\ell^{-}}\left( \eta\left( j\right) ,\text{off}\right)$
leading to a mirror-antisymmetric function of the jet rapidity
$A_{FB}^{\ell }\left( \eta\left( j\right) ,\text{off}\right)
=-A_{FB}^{\ell}\left( -\eta\left( j\right) ,\text{off}\right) $.
\ Thus the asymmetry need not vanish locally in $\eta\left( j\right) $,
but must yield zero when integrated over a symmetric interval of the
jet rapidity ($-\eta_{0}>\eta\left( j\right) >\eta_{0}$). \ The
$p_{T}$ correlation of Eq.~\ref{asympt} is such an integrated quantity
and is exhibited in Fig.~\ref{fig:ptasym_pythia2}. \ This figure
provides a comparison of $A_{FB}^{\ell^{+}}$ between the
\textsc{pythia} results with the color-flow correlation\
\textquotedblleft off\textquotedblright\ (\texttt{MSTP(67)} and
\texttt{MSTJ(50)} \textquotedblleft off\textquotedblright, the solid 
curve), \textsc{pythia} results with the color-flow correlation
\textquotedblleft on\textquotedblright\ (the dashed curve) and the 
results of the NLO matrix element calculation (the dot-dashed curve),
which we already considered in Fig. \ref{fig:asym-pT.tt1j.ll}. \ As
expected, the \textquotedblleft off\textquotedblright\ curve is
consistent with zero, while the \textquotedblleft
on\textquotedblright\ \textsc{pythia} result and the NLO matrix
element calculation exhibit comparable, negative values for
\begin{figure}[th]
\begin{center}
\includegraphics[scale=0.83]{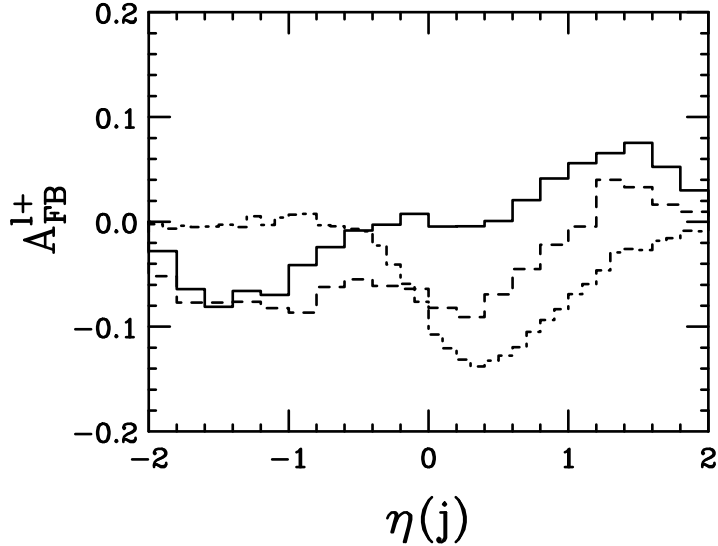} \vspace{-7mm}
\end{center}
\caption{Asymmetry $A_{FB}^{\ell^{+}}$ as a function of extra jet
pseudorapidity $\eta(j)$ with \textsc{pythia} color-flow correlations
turned \textquotedblleft off\textquotedblright\ (solid) and
\textquotedblleft on\textquotedblright\ (dashed), as compared with 
the matrix element prediction (dot-dashed).}
\label{fig:etaasym_pythia1}
\end{figure}
$A_{FB}^{\ell^{+}}$ ($A_{FB}^{\ell^{-}}$ has equal magnitude but the
opposite sign) with little $p_{T}$ dependence.

Similarly to our earlier discussion, the variation of the asymmetry as
a function of the jet rapidity is a richer subject as displayed in
Fig.~\ref{fig:etaasym_pythia1}. \ As already argued the
\textsc{pythia} \textquotedblleft off\textquotedblright\ results show 
a mirror-antisymmetric function of the jet rapidity,
$A_{FB}^{\ell}\left( \eta\left( j\right) ,\text{off}\right)
=-A_{FB}^{\ell}\left( -\eta\left( j\right) ,\text{off}\right)$, which
integrates to zero over a symmetric $\eta\left( j\right) $ interval.
\ In this case the \textquotedblleft no radiation\textquotedblright\
exclusive sample will exhibit no asymmetry. \ Note that the
\textquotedblleft off\textquotedblright\ distribution in 
Fig.~\ref{fig:etaasym_pythia1} is similar in shape to the $gg$
subprocess results in Figs.~\ref{fig:asym-y.tt1j.ll}
and~\ref{fig:asym-y.tt1j.lh}, but with the opposite sign. \ The sign
of the asymmetry correlation can be understood by noting that in the
color-flow correlation \textquotedblleft off\textquotedblright\
\textsc{pythia} calculation the asymmetry arises predominantly from
cases where the extra jet comes from final state radiation (FSR) with
the jet and the $t\bar{t}$ pair in the same hemisphere. \ This is to
be contrasted with the $gg$ subprocess NLO matrix element calculation
where the extra jet is recoiling from the $t\bar{t}$ pair and tends to
be in the opposite hemisphere. \ As indicated in
Fig.~\ref{fig:etaasym_pythia1} for the color-flow correlation
\textquotedblleft on\textquotedblright\ case (and the real matrix
element result) the asymmetry knows about the $C$ non-invariant
structure of the initial state allowing a variety of features:
$A_{FB}^{\ell^{+}}\left( \eta\left( j\right) ,\text{on}\right) \neq
A_{FB}^{\ell^{-}}\left( \eta\left( j\right) ,\text{on}\right)$,
$A_{FB}^{\ell }\left( \eta\left( j\right) ,\text{on}\right)
\neq-A_{FB}^{\ell}\left( -\eta\left( j\right) ,\text{on}\right)$, and
$A_{FB}^{\ell}\left( \eta\left( j\right) ,\text{on}\right)$ yields a
non-zero net value when integrated over the jet rapidity. \ The fact
that the net asymmetry for $\ell^{+}$ is negative (positive for
$\ell^{-}$) arises not from these symmetry considerations, but rather
the details of the correlations. The corresponding \textquotedblleft
no radiation\textquotedblright\ complement has an asymmetry of the
opposite, \textit{i.e.}, positive, sign (for $\ell^{+}$).

\begin{figure}[th!]
\begin{center}
\includegraphics[scale=0.85]{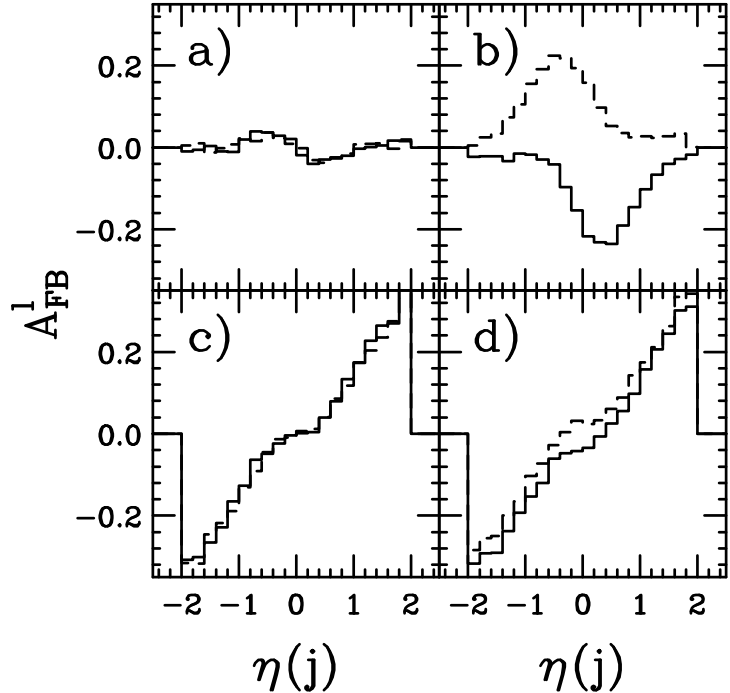} \vspace{-2mm} \vspace{-5mm}
\end{center}
\caption{Asymmetry $A_{FB}^{\ell}$ as a function of extra jet 
pseudorapidity $\eta(j)$ for $\ell^{+}$ (solid) and $\ell^{-}$
(dashed) for the following choices of \textsc{pythia} parameters: a)
with color-flow correlations turned \textquotedblleft
off\textquotedblright, ISR \textquotedblleft on\textquotedblright\ and
FSR \textquotedblleft off\textquotedblright; b) with color-flow
correlations turned \textquotedblleft on\textquotedblright, ISR
\textquotedblleft on\textquotedblright\ and FSR \textquotedblleft
off\textquotedblright; c) with color-flow correlations turned
\textquotedblleft off\textquotedblright, ISR \textquotedblleft
off\textquotedblright\ and FSR \textquotedblleft on\textquotedblright;
d) with color-flow correlations turned \textquotedblleft
on\textquotedblright, ISR \textquotedblleft off\textquotedblright\ and
FSR \textquotedblleft on\textquotedblright.}
\label{fig:pythia2}
\end{figure}

We can analyze the \textsc{pythia} calculations in somewhat more
detail by separately turning \textquotedblleft on\textquotedblright\
and \textquotedblleft off\textquotedblright\ color-flow correlations,
initial state radiation (ISR, as defined in \textsc{pythia}) and final
state radiation (FSR, as defined in \textsc{pythia}). \ (Note that,
due to interference effects there is no directly analogous analysis
possible for the matrix element calculation.) \ The corresponding
curves are exhibited in Fig.~\ref{fig:pythia2} for both $\ell^{+}$
(solid) and $\ell^{-}$ (dashed). \ (The case with all parameters set
to \textquotedblleft on\textquotedblright\ is the dashed cure in
Fig.~\ref{fig:etaasym_pythia1} and the case with all radiation
\textquotedblleft on\textquotedblright\ but color-flow correlations
\textquotedblleft off\textquotedblright\ is the solid curve in 
Fig.~\ref{fig:etaasym_pythia1}.) \ As suggested earlier, we see in
Fig.~\ref{fig:pythia2} a) and c) that with no color-flow correlations
$A_{FB}^{\ell^{+}}\left( \eta\left( j\right) ,\text{off}\right)
=A_{FB}^{\ell ^{-}}\left( \eta\left( j\right) ,\text{off}\right)$.
\ On the other hand in Fig.~\ref{fig:pythia2} b) and d) with color-flow
correlations present we have $A_{FB}^{\ell^{+}}\left( \eta\left(
j\right) ,\text{on}\right) \neq A_{FB}^{\ell^{-}}\left( \eta\left(
j\right) ,\text{on}\right) $. \ From Fig.~\ref{fig:pythia2} a) we
learn that ISR without color-flow correlations yields just a small
(mirror-antisymmetric) asymmetry with the radiation in the opposite
direction from the $t\bar{t}$ pair, while the addition of color-flow
correlations in b) yields a much larger asymmetry with the $t$ quark
recoiling from (moving opposite to) the radiation (with the $\bar{t}$
approximately at rest) for forward radiation and the $\bar{t}$ moving
forward (with the $t$ approximately at rest) for backward radiation.
\ It is this structure that is qualitatively similar to the $q\bar{q}$
subprocess matrix element result that dominates in
Fig.~\ref{fig:asym-y.tt1j.ll}. \ Comparing Fig.~\ref{fig:pythia2} c)
and d) we see that the FSR contribution corresponds to the $t\bar{t}$
pair and the radiation all tending to move in the same direction, and
that this mirror-antisymmetric structure is only slightly modified
when color-flow correlations are present. \ Further, it is the FSR
contribution that defines the qualitative shape of the combined result
with no color-flow correlations present (the solid curve in
Fig.~\ref{fig:etaasym_pythia1}). \ The combined result including
color-flow correlations, the dashed curve in
Fig.~\ref{fig:etaasym_pythia1}, has a shape that interpolates between
Fig.~\ref{fig:pythia2} c) and Fig.~\ref{fig:pythia2} d). \ However,
since the FSR contribution in d) is close to mirror-antisymmetric, it
is the ISR contribution in b) that dominates any symmetric integral
over this distribution.

The lesson here is that, even though \textsc{pythia} does not (cannot)
exhibit the inclusive forward-backward asymmetry characteristic of NLO
QCD, \textsc{pythia} does include parameters that can be set to
produce events with much of the correlated asymmetry structure of NLO
QCD. \ However, the agreement is only qualitative. \ \textit{In
detail} the asymmetries in the \textsc{pythia} event samples have
neither the same magnitude nor the same dependence on the extra
radiation as the matrix element prediction, \textit{i.e.}, the dashed
and dot-dashed curves in Figs.~\ref{fig:ptasym_pythia2}
and~\ref{fig:etaasym_pythia1} are distinct. At least one underlying
difference is that \textsc{pythia} includes FSR from the b quark in
the top decay, which we do not include in the matrix element
calculations. \ To the extent that the color-flow structure in
\textsc{pythia} correctly reflects coherence effects in the higher
order showering corrections, beyond NLO, the asymmetries exhibited by
\textsc{pythia} are suggestive of what higher-order perturbative
analyses will yield.

We have not studied \textsc{herwig}~\cite{herwig} extensively except
to confirm that the inclusive asymmetry is not present. It may or may
not have correlated asymmetries similar to those observed with
\textsc{pythia} arising from color-flow related constraints on extra
radiation from showers. \ We have performed no analogous analyses with
\textsc{sherpa}~\cite{sherpa}. However, future studies using these
generators should be aware that these correlated asymmetries may be
present. \ In any case, it is clear that the experimental study of
asymmetries, both inclusive and exclusive, can yield information about
the coherence structure of QCD showers. \ To the extent that top quark
physics is a background in studies of physics beyond the Standard
Model, we must come to understand quantitatively these correlations
and asymmetries in Monte Carlo simulations. \

%%%%%%%%%%%%%%%%%%%%%%%%%%%%%%%%%%%%%%%%%%%%%%%%%%%%%%%%%%%%%%%%%%%%%%%%

\section{Conclusions}

\label{sec:conc}

Top quark pair production at a $p\bar{p}$ collider exhibits a forward-backward
asymmetry due to higher-order short-distance QCD effects. This has been known
for almost two decades, but only at an abstract theoretical level.
Contributions to the asymmetry arise both from virtual corrections to
$q\bar{q}\to t\bar{t}$, due to interference between the color-singlet
box (4-pt.) and Born terms, and from real gluon emission in the same
subprocess, due to interference of emission between the initial and final
states (although one cannot think of it physically so simply, because those
and other diagrams for emission are gauge-related). The two contributions are
opposite in sign and of similar magnitude, leading to a partial cancellation
which leaves a smaller asymmetry in the inclusive sample of the same sign as
the virtual piece.

We have explored the asymmetry here with an eye toward experimental
measurement. There are several major new points. First, experimentally the
most reliable avenue for measuring the asymmetry is likely via the charged
leptons, where both the direction and the sign of the charge can be determined
with confidence. Unfortunately, we cannot test this approach in detail for an
inclusive $t\bar{t}$ sample without a (currently unavailable) full NLO
calculation including the $t$ decays. \ For the $t\bar{t}j$ sample (requiring
extra radiation), where the asymmetry appears at lowest non-trivial order, our
results do confirm that the asymmetry at the top quark level is reflected in
the top quark decay products, although with different magnitude. This change
in the magnitude is due to the top quarks' spins leading to decay products
being emitted preferentially in certain directions, which can move them from
one hemisphere to another. Secondly, we have pointed out that the asymmetry of
the leptons \ in the $t\bar{t}j$ sample is correlated with the angular
distribution of additional jets in the event, but not their transverse
momentum. \ Unfortunately, the real-emission asymmetries, which we term the
$t\bar{t}j$ and $t\bar{t}jj$ components, are presently known only at LO. These
last aspects of the short-distance asymmetry were not previously noticed in
the literature. \ Finally we have studied the likely numbers of events at the
Tevatron for the fully inclusive $t\bar{t}$ sample, where we assume the
leptonic asymmetry is equal to the top-quark asymmetry, the radiation-required
$t\bar{t}j$ sample, and the jet-veto-based exclusive $t\bar{t}0j$ sample. \ In
all cases statistics will be an issue at the Tevatron, requiring a multiple
channel analysis with (hopefully) outstanding accelerator and detector performance.

We have also shown that some models for color-flow correlations in QCD showers
can yield correlated asymmetries as well, as exemplified in the parton shower
Monte Carlo \textsc{pythia}\cite{pythia}. While the lepton asymmetry always
vanishes in the totally inclusive \textsc{pythia}-generated sample, when the
relevant shower parameters are set (to the default values) to exhibit a
specific model of color-flow structure\ in the parton showers, a negative top
quark asymmetry arises in the $t\bar{t}j$ component with a positive asymmetry
in the corresponding exclusive $0j$ component. Both the overall magnitude of
the asymmetry and the differential asymmetry with respect to the additional
hard jet angular distribution are qualitatively similar to but different in
detail from those predicted by the perturbative NLO results.

We again emphasize that we have tried to explore only the general structure of
the $t\bar{t}$ asymmetry signal and that, due to a number of limitations in
the current theoretical tools, our results should not be regarded as precise
predictions. We review here the major sources of those
uncertainties/questions, which should be considered as areas that theorists
need to address in the near future:

\begin{itemize}
\item[$\cdot$] How does the inclusive asymmetry change once kinematic cuts are
imposed on the decay products? No NLO calculation exists to address this question.

\item[$\cdot$] What is the overall $t\bar{t}j$ normalization, which is known
presently only at LO? Additionally, we need to evaluate the NLO rate including
decays so that cuts may be imposed.

\item[$\cdot$] What is the effect of radiative top quark decays on
experimental efficiencies for correctly selecting the extra jet in the
inclusive production-radiation $t\bar{t}j$ sample? We need this to be able to
isolate the maximally-asymmetric region in $\eta(j)$.

\item[$\cdot$] We need to properly merge the $t\bar{t}j$ matrix elements into
the parton-shower Monte Carlo environment in order to be able to make complete predictions.
\end{itemize}

While the experimental measurement of the asymmetries described here will
likely be a challenge at the Tevatron due to limited statistics, it is still
an extremely important goal. Since top quarks may well have a unique
connection to new physics, and since they play a large role as a background in
many new physics searches, Tevatron experimentalists should strive to
determine top quark properties as accurately as possible. Further, the
analysis of the correlation between the asymmetry and extra radiation may
offer a nearly unique opportunity to elucidate detailed properties of
long-distance, all-orders QCD, \textit{i.e.}, the color-flow\ structure of the
parton shower. While the measurement of $A_{FB}^{\ell}$ is unlikely to be a
direct probe for new physics, unless the new physics contributions to the
asymmetry are extremely large or of opposite sign to the SM, this measurement
still provides a good example of the subtle behavior of SM particles in
unexpected places. We should understand the SM as well as possible, from both
theory and data, before moving on to see what lies beyond.

%%%%%%%%%%%%%%%%%%%%%%%%%%%%%%%%%%%%%%%%%%%%%%%%%%%%%%%%%%%%%%%%%%%%%%%%

\begin{acknowledgments}
This research was supported in part by the U.S. Department of Energy under
grant Nos. DE-FG02-91ER40685 and DE-FG02-96ER40956. D.R. would like to thank
Regina Demina and Paul Tipton for advice on capabilities of the D\O \ and CDF
detectors in Run~II, while S.D.E. and M.T.B. thank Gordon Watts, Henry
Lubatti, Toby Burnett and Matt Strassler for many helpful conversations.
\end{acknowledgments}

%%%%%%%%%%%%%%%%%%%%%%%%%%%%%%%%%%%%%%%%%%%%%%%%%%%%%%%%%%%%%%%%%%%%%%%%

%%%%%%%%%%%%%%%%%%%%%%%%%%%%%%%%%%%%%%%%%%%%%%%%%%%%%%%%%%%%%%%%%%%%%%%%
%%% References
%%%%%%%%%%%%%%%%%%%%%%%%%%%%%%%%%%%%%%%%%%%%%%%%%%%%%%%%%%%%%%%%%%%%%%%%

\end{document}